\documentclass[12pt,preprint]{aastex}

\slugcomment{To appear in the June 2004 issue of {\it The Astronomical Journal}}

\shorttitle{$L'M'$ Photometry of Ultracool Dwarfs}
\shortauthors{Golimowski et al.}
 
\begin{document}
 
\title{$L'$ and $M'$ Photometry of Ultracool Dwarfs}

\author{D.~A.\ Golimowski,\altaffilmark{1}
S.~K.\ Leggett,\altaffilmark{2}
M.~S.\ Marley,\altaffilmark{3} 
X.\ Fan,\altaffilmark{4}
T.~R.\ Geballe,\altaffilmark{5}
G.~R.\ Knapp,\altaffilmark{6}
F.~J.\ Vrba,\altaffilmark{7}
A.~A.\ Henden,\altaffilmark{7}
C.~B.\ Luginbuhl,\altaffilmark{7}
H.~H.\ Guetter,\altaffilmark{7}
J.~A.\ Munn,\altaffilmark{7}
B.\ Canzian,\altaffilmark{7}
W.\ Zheng,\altaffilmark{1}
Z.~I.\ Tsvetanov,\altaffilmark{8}
K.\ Chiu,\altaffilmark{1}
K.\ Glazebrook,\altaffilmark{1}
E.~A.\ Hoversten,\altaffilmark{1}\\
D.~P.\ Schneider,\altaffilmark{9,11}
and
J.\ Brinkmann\altaffilmark{10,11}
}

\altaffiltext{1}{ 
Department of Physics and Astronomy, 
The Johns Hopkins University, 
3400 North Charles Street, 
Baltimore, MD 21218-2686
}
\altaffiltext{2}{ 
United Kingdom Infrared Telescope, 
Joint Astronomy Centre, 
660 North A'ohoku Place, 
Hilo, HI 96720
}
{
\altaffiltext{3}{ 
NASA--Ames Research Center, 
Mail Stop 245-3,
Moffett Field, CA 94035 
}
\altaffiltext{4}{ 
Steward Observatory,
The University of Arizona,
Tucson, AZ 85721-0065
}
\altaffiltext{5}{ 
Gemini Observatory, 
670 North A'ohoku Place,
Hilo, HI 96720
}
\altaffiltext{6}{ 
Princeton University Observatory, 
Princeton, NJ 08544
}
\altaffiltext{7}{ 
United States Naval Observatory, 
Flagstaff Station, 
P.O. Box 1149, 
Flagstaff, AZ 86002-1149
}
\altaffiltext{8}{ 
National Aeronautics and Space Administration,
300 E Street SW, 
Washington, DC 20546-0001
}
\altaffiltext{9}{
Department of Astronomy and Astrophysics,
Pennsylvania State University,
525 Davey Laboratory,
University Park, PA 16802
}
\altaffiltext{10}{
Apache Point Observatory,
2001 Apache Point Road,
P.O.\ Box 59,
Sunspot, NM 88349-0059
}
\altaffiltext{11}{
Sloan Digital Sky Survey builder.
}

\begin{abstract} 
We have compiled $L'$ (3.4--$4.1~\mu$m) and $M'$ (4.6--$4.8~\mu$m) photometry of 63 single and binary M, L,
and T dwarfs obtained at the United Kingdom Infrared Telescope using the Mauna Kea Observatory (MKO) filter
set.  This compilation includes new $L'$ measurements of 8 L dwarfs and 13 T dwarfs and new $M'$ measurements
of 7 L dwarfs, 5 T dwarfs, and the M1 dwarf Gl~229A.  These new data increase by factors of 0.6 and 1.6, 
respectively, the numbers of ultracool dwarfs ($T_{\rm eff} \lesssim 2400$~K) for which $L'$ and $M'$ 
measurements have been reported.  We compute $L_{\rm bol}$, BC$_K$, and $T_{\rm eff}$ for 42 dwarfs whose 
flux-calibrated $JHK$ spectra, $L'$ photometry, and trigonometric parallaxes are available, and we estimate 
these quantities for 9 other dwarfs whose parallaxes and flux-calibrated spectra have been obtained.  BC$_K$
is a well-behaved function of near-infrared spectral type with a dispersion of $\sim 0.1$~mag for types M6--T5;
it is significantly more scattered for types T5--T9.  $T_{\rm eff}$ declines steeply and monotonically for 
types M6--L7 and T4--T9, but is nearly constant at $\sim 1450$~K for types L7--T4 with assumed ages of 
$\sim 3$~Gyr.  This constant $T_{\rm eff}$ is evidenced by nearly unchanging values of $L'$--$M'$ between types
L6 and T3.  It also supports recent models that attribute the changing near-infrared luminosities and spectral 
features across the L--T transition to the rapid migration, disruption, and/or thinning of condensate clouds 
over a narrow range of $T_{\rm eff}$.  The $L'$ and $M'$ luminosities of early-T dwarfs do not exhibit the 
pronounced humps or inflections previously noted in the $I$ through $K$ bands, but insufficient data exist for 
types L6--T5 to assert that $M_{L'}$ and $M_{M'}$ are strictly monotonic within this range of types.  We compare 
the observed $K$, $L'$, and $M'$ luminosities of L and T dwarfs in our sample with those predicted by 
precipitating-cloud and cloud-free models for varying surface gravities and sedimentation efficiencies.  The models 
indicate that the L3--T4.5 dwarfs generally have higher gravities (log~$g = 5.0$--5.5) than the T6--T9 dwarfs 
(log~$g = 4.5$--5.0).  The predicted $M'$ luminosities of late-T dwarfs are 1.5--2.5 times larger than those derived 
empirically for the late-T dwarfs in our sample.  This discrepancy is attributed to absorption at 4.5--$4.9~\mu$m
by CO, which is not expected under the condition of thermochemical equilibrium assumed in the models.  Our 
photometry and bolometric calculations indicate that the L3 dwarf Kelu-1 and the T0 dwarf 
SDSS~J042348.57--041403.5 are probable binary systems.  We compute log$(L_{\rm bol}/L_{\odot}) = -5.73 \pm 0.05$
and $T_{\rm eff} = 600$--750~K for the T9 dwarf 2MASSI~J0415195--093506, which supplants Gl~570D as the least 
luminous and coolest brown dwarf presently known.
\end{abstract}
 
\section{Introduction}

The number of known ultracool dwarfs -- dwarfs with effective temperatures $T_{\rm eff} \lesssim 2400$~K -- has 
grown dramatically over the last seven years, primarily because of the sizes and depths of the DEep Near-Infrared
Survey of the Southern Sky \citep[DENIS;][]{epc97}, the Two-Micron All Sky Survey \citep[2MASS;][]{skr97}, and 
the Sloan Digital Sky Survey \citep[SDSS;][]{yor00}.  These surveys, and others of lesser scope, have revealed 
ultracool dwarfs in numbers sufficient to populate a distribution of temperatures ranging from the coolest 
hydrogen-burning stars \citep[$1550 \lesssim T_{\rm eff} \lesssim 1750$~K;][]{bur93,cha00} to the coolest known 
brown dwarf ($600 \lesssim T_{\rm eff} \lesssim 750$~K; this paper).  Consequently, two new spectral types, L and 
T, have been defined in order to classify dwarfs cooler than spectral type M \citep[hereafter G02]
{kir99b,mar99b,bur02a,geb02}.  The spectra of L dwarfs are characterized by absorption from neutral alkali metals 
(e.g., K, Na, Cs, and Rb) and metallic hydrides (e.g., FeH and CrH) at red wavelengths and by absorption from CO 
and H$_2$O at near-infrared wavelengths.  In contrast, the optical spectra of T dwarfs are dominated by 
pressure-broadened K~{\small I} and Na~{\small I} absorption lines, and their near-infrared spectra are sculpted 
by broad absorption bands of CH$_4$ and H$_2$O and collisionally-induced absorption (CIA) by H$_2$. 

Most observational studies of ultracool dwarfs concern the spectral region 0.6--2.5~$\mu$m.  This region is 
favored because the flux distributions of these objects peak around 1.2~$\mu$m and because the spectral 
sensitivities of modern photoelectronic detectors coincide with the relatively dark and transparent atmospheric 
windows in this region.  Recently, attention has been given to the photometry of ultracool dwarfs longward of the
$K$ bandpass. This attention has been motivated partly by the need to better constrain the bolometric luminosities
of ultracool dwarfs and by the prospects for observing even cooler brown dwarfs or planets at these wavelengths 
with space-based infrared telescopes.  The intrinsic faintness of L and T dwarfs and the increasing brightness 
and variability of the sky make ground-based observations of these objects in the $L$ (2.5--4.0~$\mu$m) and $M$ 
(4.1--5.5~$\mu$m) bandpasses difficult and time-consuming.  Although $L$-band photometry of L and T dwarfs is 
extensive \citep[hereafter L02]{jon96,leg98,tok99,ste01,rei02,leg02a}, $M$-band photometry has been published 
for only six of these objects (\citealt{mat96,rei02}; L02).

Spectroscopic studies of brown dwarfs in the $L$ or $M$ bandpasses have so far been limited to three L dwarfs 
and the archetypal T dwarf, Gl~229B.  The $L$ bandpass contains the Q-branch of the fundamental absorption band 
of CH$_4$, which is situated near 3.3~$\mu$m.  This absorption band appears as early as spectral type L5 
\citep[$T_{\rm eff} \approx 1700$~K;][]{nol00} and is deep and broad in the spectrum of the T6 dwarf Gl~229B 
\citep[$T_{\rm eff} \approx 900$~K;][]{opp98}.  The $M$-band spectrum of Gl~229B features a broad but shallow 
absorption trough from 4.5~$\mu$m to 4.9~$\mu$m and a narrow peak at 4.67~$\mu$m, both of which are attributed 
to the 1-0 vibration-rotation band of CO \citep{nol97,opp98}.  These features reveal a CO abundance that is over
1000 times larger than expected under conditions of CO$\leftrightarrow$CH$_4$ thermochemical equilibrium, 
indicating that CO is rapidly transported outward from warmer, CO-rich layers of the atmosphere 
\citep{feg96,gri99,sau00,sau03}.  The $M$-band fluxes of two other T dwarfs, SDSS~J125453.90--012247.4 and
2MASS~J05591914--1404488, are reportedly well below the levels expected for CO$\leftrightarrow$CH$_4$ equilibrium, 
which suggests that vertical mixing of CO within the atmospheres of T dwarfs is common \citep[L02;][]{sau03}. 

In this paper, we present new 3.4--$4.1~\mu$m and 4.6--$4.8~\mu$m photometry of ultracool dwarfs obtained with the 
United Kingdom Infrared Telescope (UKIRT) using the Mauna Kea Observatory (MKO) $L'$ and $M'$ filters.  These data
increase by factors of 0.6 and 1.6, respectively, the numbers of ultracool dwarfs for which MKO $L'$ and $M'$ 
measurements have been reported.  We examine the near-infrared colors and magnitudes of these dwarfs as functions 
of spectral type.  Using recently published trigonometric parallaxes, we show color--magnitude diagrams in the MKO
$K$, $L'$, and $M'$ bandpasses, and we determine the bolometric luminosities and effective temperatures of ultracool
dwarfs.  We compare these results with the predictions of recent atmospheric models that consider the effects of 
cloud sedimentation on the broadband spectra of these objects.  Finally, we consider the effects of nonequilibrium 
CO$\leftrightarrow$CH$_4$ chemistry on the $M'$ luminosities of ultracool dwarfs and on direct searches for even 
cooler objects at wavelengths around $5~\mu$m. 

\section{The Sample}

The sample of objects under study comprises 63 single and binary M, L, and T dwarfs for which MKO $L'M'$ 
photometry has been presented by L02, \citet{leg02b}, \citet{rei02}, or in this paper.  Although our study 
concerns ultracool dwarfs (spectral types M7 and later), we include in our sample some early-M dwarfs to 
establish a connection with the cool end of the classical main sequence.  Counting only the single dwarfs and 
the primary components of close binaries, our sample numbers 15~M dwarfs, 28~L dwarfs, and 20~T dwarfs.  Table~1
lists the names, multiplicities, spectral types, trigonometric parallaxes, and distance moduli of the dwarfs in 
our sample, as well as published references for those characteristics.  The parallaxes and distance moduli listed
in columns~3 and 4 are based upon the weighted means of the parallax measurements referenced in column~6.  The 
names of the dwarfs are the full designations assigned to them by the catalogues or surveys of their origin, 
using (where possible) the most current naming protocols for those sources.  Henceforth, we abbreviate the names 
of the single and binary dwarfs detected by DENIS, 2MASS, and SDSS using their survey acronyms, followed by the 
first four digits of both their Julian right ascensions and declinations.  These abbreviated forms are preferred 
by the International Astronomical Union (IAU).  

The spectral types of all but two of the L and T dwarfs listed in Table~1 are derived from their $J$-, $H$-, and 
$K$-band spectra using the near-infrared spectral classification scheme of G02.  The types listed for the primary
components of close-binary systems are derived from the composite spectra of the binaries using this classification
scheme.  The types listed for the secondary components are either previously published estimates or new estimates 
based on published luminosities and the relationship between luminosity and spectral type presented by L02.  The 
classification scheme of G02 employs four indices that measure the strengths of H$_2$O and CH$_4$ absorption bands
between $1.1~\mu$m and $2.3~\mu$m.  The monotonic variation of these indices through the L and T sequences permits
classification of these dwarfs with a typical uncertainty of one-half spectral subtype.  The H$_2$O $1.5~\mu$m 
index and a fifth index measuring the slope of the red continuum flux are well suited for classifying early-L dwarfs.
These indices yield early-L types that are consistent with types obtained from the optical classification schemes of
\citet{kir99b} and \citet{mar99b}.  However, discrepancies between the optical and G02 schemes as large as 2.5 
subtypes occur for mid- to late-L dwarfs, which suggests that the optical and near-infrared indices are unequally 
affected by the changing opacity of condensate clouds as $T_{\rm eff}$ decreases \citep{ste03}.  The spectral types 
of T dwarfs obtained from the independent near-infrared classification schemes of G02 and \citet{bur02a} usually match 
within one-half subtype.

We have obtained new MKO $L'$ photometry for 21 dwarfs in our sample (8 L dwarfs and 13 T dwarfs) and new MKO $M'$ 
photometry for 13 dwarfs in our sample (1 M dwarf, 7 L dwarfs, and 5 T dwarfs).  These new data increase to 57 and 
21 the numbers of ultracool dwarfs that have been measured photometrically in the MKO $L'$ and $M'$ bandpasses, 
respectively \citep[L02;][]{leg02b,rei02}.  This group of ultracool dwarfs is the largest so far measured in any 
single $L$- and $M$-band photometric system.  \citet{ste01} obtained $L$-band photometry of 23 ultracool dwarfs 
using the $L'$ and $L_s$ filters installed in the Near Infrared Camera (NIRC) at the W.~M. Keck Observatory on 
Mauna Kea, Hawaii.  Because no photometric transformations between the Keck $L'$ and $L_s$ bandpasses and the MKO 
$L'$ bandpass exist yet for ultracool dwarfs, we exclude these 23 measurements from our present analysis.  

\section{Observations and Data Reduction}

The new $L'$ and $M'$ photometric data were obtained between 2001 November and 2003 November using the 3.8~m 
UKIRT on Mauna Kea, Hawaii.  Data obtained before 2002 September~1 were recorded with UKIRT's 1--5~$\mu$m InfraRed 
Camera \citep[IRCAM;][]{pux94}; data obtained thereafter were recorded with the new 1--5~$\mu$m UKIRT Imager 
Spectrometer \citep[UIST;][]{ram00}.  IRCAM features a $256 \times 256$ array of $30~\mu$m InSb pixels and optics 
that yield a pixel scale of 0\farcs081~pixel$^{-1}$ and a field of view of 20\farcs7~$\times$~20\farcs7.  UIST 
features a $1024 \times 1024$ ALADDIN array of $27~\mu$m InSb pixels and selectable optics in its imaging mode 
that yield pixel scales of 0\farcs06~pixel$^{-1}$ and 0\farcs12~pixel$^{-1}$.  The former scale and a $512 \times 
512$ subarray readout were used to increase the efficiency of our observations.  This configuration provided a 
field of view of 30\farcs7~$\times$~30\farcs7.  Both imagers are equipped with broadband filters spanning the 
range 1.15--4.9$~\mu$m, including the $L'$ and $M'$ filters of the MKO photometric system \citep{sim02,tok02}.
Descriptions of these filters and the differences between commonly used $L$- and $M$-bandpasses have been presented
by L02.

To investigate possible differences between the instrumental $L'$ and $M'$ magnitudes of IRCAM and UIST, we 
synthesized the $L'$ and $M'$ magnitudes of Gl~229B by convolving its 3.0--$4.2~\mu$m and 4.5--$5.1~\mu$m spectra 
\citep{nol97,opp98} with the measured transmission and reflection profiles of the imagers' optics and detectors.
A 5\% dip in the 2.7--$3.5~\mu$m transmission of UIST's lenses produces a value of $L'$ that is 0.015~mag larger 
than that computed for IRCAM.  No other instrumental features affect the $L'$ magnitudes significantly.  The $M'$ 
magnitudes computed for each imager are nearly identical.  The differences between the pairs of synthetic magnitudes 
are much less than the random errors associated with actual $L'$ and $M'$ magnitudes of ultracool dwarfs obtained 
with either imager (Table~2).  Thus, our limited investigation indicates that $L'$- and $M'$-band measurements 
recorded with IRCAM and UIST are compatible with the MKO $L'$ and $M'$ photometric system.

Table~2 lists the dates and instruments of observation and the calibrated magnitudes of the 28 dwarfs in our sample 
for which new $L'$ or $M'$ photometric data were obtained.  All data were recorded during photometric (dry and 
cloudless) conditions and subarcsecond seeing.  The techniques of recording and reducing the data from both imagers 
mimicked those of previous IRCAM observations of ultracool dwarfs (L02).  Because of the bright sky background, the 
reduced images comprised scores of short-exposure, co-added frames.  Typically, each $L'$ exposure consisted of 100
co-added exposures of 0.2~s, and each $M'$ exposure comprised 75 co-added exposures of 0.12~s.  The telescope was 
offset slightly between frames.  Adjacent pairs of frames were subtracted to remove the rapidly varying background 
signal, and every four pairs of differenced images were combined and divided by a flat field.  This process was 
repeated until the desired ratio of signal to noise (S/N) was achieved.  With IRCAM, S/N~$\approx 20$ was achieved 
in 1--30~min for targets having $L' \approx 11$--13.  Likewise, S/N~$\approx 10$ was reached in about 1.3~hr for 
targets having $M' \approx 12$.  With UIST, S/N~$\approx 15$ was achieved in 1~hr for targets having $L' \approx 13$,
and S/N~$\approx 13$ was achieved in 1~hr for targets having $M' \approx 11$.  All data were calibrated using UKIRT
standard stars observed through the MKO $L'$ and $M'$ filters \citep{leg03}.

\section{Results and Analysis}

Table~3 lists the new and previously published MKO $KL'M'$ photometry for our sample of M, L, and T dwarfs.  All 
measurements come from L02 or this paper unless otherwise noted.  Absolute $L'$ magnitudes are also listed for 
objects whose trigonometric parallaxes have been published.  The values of $M_{L'}$ for five T dwarfs are based on
the weighted means of absolute parallaxes reported by \citet[hereafter D02]{dah02} and \citet[hereafter V04]{vrb04}
and relative parallaxes reported by \citet{tin03}.  The $M_{L'}$ for another T dwarf, 2MASS~J1534--2952AB, is based
solely upon a relative parallax.  \citet{tin03} estimate that the corrections from relative to absolute parallaxes
are less than 0\farcs001 for their astrometric fields, so the systematic differences between values of $M_{L'}$ 
derived separately from absolute and relative parallaxes are probably within 0.06~mag for the dwarfs in our sample. 

\subsection{$K$--$L'$ and $L'$--$M'$ colors}

Figure~\ref{Golimowski.fig1} shows the variations of $K$--$L'$ and $L'$--$M'$ with spectral type for the dwarfs in our 
sample.  The ordinate axes of each diagram have the same incremental scales so that the relative changes in each 
color may be compared directly.  Figure~\ref{Golimowski.fig2} is a diagram of $K$--$L'$ versus $L'$--$M'$ for dwarfs measured 
in all three bandpasses.  In these figures, M dwarfs are denoted by circles, L dwarfs by triangles, and T dwarfs by 
squares.  Points representing close-binary systems are surrounded by open circles to distinguish them from single or 
widely separated dwarfs.  These representations are maintained throughout this paper. 

The diagram of $K$--$L'$ versus spectral type is a more populated version of the similar diagram shown by L02.  
$K$--$L'$ generally increases monotonically through the spectral sequence, but the rate of increase changes 
significantly over the range of spectral types shown.  The reddening between types M1 and L0 is approximately linear 
with a dispersion of $\sim 0.06$~mag, but $K$--$L'$ increases nonlinearly through the L sequence with a dispersion 
of $\sim 0.15$~mag.  This scatter has been attributed to the strong and varying effects of condensate clouds on the 
emergent $K$- and $L$-band fluxes over the associated range of $T_{\rm eff}$ (\citealt{ack01}; L02).  Variations in 
surface gravity among L dwarfs may also contribute to this scatter (see \S5).  The nearly unchanging values of 
$K$--$L'$ between types L6 and T5 were noted by L02, who attributed this behavior to the balanced effects of 
increasing CH$_4$ absorption at 2.2--$2.4~\mu$m and 3.3--$3.7~\mu$m, the latter of which extends into the blue half 
of the $L'$ band.  Similar behavior can be inferred from the Keck $K$--$L'$ measurements of \citet{ste01} despite a 
lack of data for types L8.5--T0.5.  As we discuss in \S4.3, this behavior may also reflect the redistribution of 
flux caused by the settling of condensate clouds in the photosphere.  The rapid increase of $K$--$L'$ beyond type 
T5 may be caused by increasing H$_2$ CIA in the $K$ band, saturation (or, alternatively, weakening) of the CH$_4$ 
absorption band at 3.3--$3.7~\mu$m, or both.  The L9 dwarf with the anomalously blue $K$--$L' = 1.20$ is 
SDSS~J0805+4812, and the T6 dwarf with the anomalously red $K$--$L' = 3.05$ is 2MASS~J0937+2931.  We discuss these
dwarfs in \S4.5.

$L'$--$M'$ decreases slowly between spectral types M5 and L6, probably because of strengthening CO absorption at 
4.5--$4.9~\mu$m \citep{tsu95a,tsu95b,rei02}.  The color is nearly constant between types L6 and T3, which suggests 
that the varying absorptions by CH$_4$ at 3.3--$3.7~\mu$m and CO at 4.5--$4.9~\mu$m have balanced effects on the 
integrated $L'$ and $M'$ fluxes, or that $T_{\rm eff}$ changes little within this range of spectral types.  While 
the former condition may be true, the results presented in \S4.3 indicate that the latter condition is certainly true.
Beyond type T5, $L'$--$M'$ rises steeply, i.e, it becomes significantly redder.  This reddening cannot be definitively
explained without a representative collection of $L$- and $M$-band spectra.  It may be caused by dissipating CO 
absorption at 4.5--$4.9~\mu$m, or it may simply reflect a Wien-like shift of the spectral energy distribution as 
$T_{\rm eff}$ decreases.

\subsection{Color--magnitude and magnitude--spectral type relations}

Figures~\ref{Golimowski.fig3}a and \ref{Golimowski.fig3}b are color--magnitude diagrams (CMDs) of $M_K$ versus $K$--$L'$ and \hbox{$K$--$M'$,}
respectively, for the dwarfs in our sample.  Figure~\ref{Golimowski.fig3}a is similar to the $M_K$ versus $K$--$L'$ diagram 
presented by L02, but it shows many more data, especially in the T dwarf domain $1.5 < K$--$L' < 2.5$.  
Figure~\ref{Golimowski.fig3}b greatly extends the equivalent $M_{M'}$ versus $K$--$M'$ diagram of \citet{rei02} by including 
L and T dwarfs with $0.7 \lesssim K$--$M' \lesssim 3.7$.  The bright end of each diagram represents the M1 dwarf 
Gl~229A; the faint end represents the T9 dwarf 2MASS~J0415--0935.  We cannot explain in detail the characteristics 
of these CMDs without a comprehensive set of 3.4--$4.8~\mu$m spectra, but their basic appearances are significant.  
The anomalous 2MASS~J0937+2931 notwithstanding (see \S4.5), both diagrams are monotonic throughout the M, L, and T 
spectral classes.  Such monotonicity, combined with a wide range of color values, is rare among CMDs constructed from 
combinations of optical and near-infrared bandpasses.  For example, the reversals of $J$--$H$ and $H$--$K$ caused by 
strengthening CH$_4$ absorption at 1.6--$1.8~\mu$m and 2.2--$2.4~\mu$m \citep[G02;][]{bur02a} cause degeneracies in 
the $JHK$-based CMDs of M and early-T dwarfs (L02; D02; \citealt{tin03}; V04; \citealt{kna04}, hereafter K04).  
Combinations of $I$-, $J$-, and $K$-band measurements produce CMDs that are similarly degenerate over a wide range 
of spectral types \citep{tin03}.  CMDs constructed exclusively from SDSS $i$ and $z$ photometry are monotonic for $i$--$z 
\lesssim 5.2$, but SDSS measurements of late-T dwarfs with known distances are presently lacking (\citealt{haw02}; 
K04).  Consequently, Figures~\ref{Golimowski.fig3}a and \ref{Golimowski.fig3}b are the only CMDs with sufficient range and resolution to 
enable reliable estimates of photometric parallaxes throughout the presently defined L and T sequences.

We have derived parametric expressions for each CMD by computing least-squares fits of polynomials to their 
respective data.  Because the intrinsic scatter of the data exceeds the photometric and astrometric measurement 
errors, the data were not weighted.  Data representing close-binary systems and 2MASS~J0937+2931 were excluded 
from the fits.  The curves in Figures~\ref{Golimowski.fig3}a and \ref{Golimowski.fig3}b are low-order polynomials that yield the optimum
$\chi^2$ statistic for the selected range of data.  The coefficients and residual statistics of the fits are 
listed in Table~4.  The fits are intended primarily for estimating the luminosities and distances of individual 
dwarfs from MKO $KL'M'$ photometry.   Because the data represent dwarfs of unknown multiplicity and age, the
fits can be used to derive only a provisional luminosity function for ultracool dwarfs.

Figures~\ref{Golimowski.fig4}a and \ref{Golimowski.fig4}b are diagrams of $M_{L'}$ and $M_{M'}$ as functions of spectral type.  
These diagrams complement the diagrams presented by \citet{tin03} and K04 for MKO $J$ and $K$ photometry reported
by L02 and K04.\footnote{\citet{tin03} incorrectly described L02's $J$ and $K$ data as having been measured on 
the old UKIRT photometric system instead of the proper MKO system.  Also, \citet{tin03} employed L types 
derived from the optical classification scheme of \citet{kir99b} rather than the near-infrared scheme of G02 used
by us.  The two schemes often yield discordant mid-L to early-T types, so systematic inconsistencies between our
Figure~\ref{Golimowski.fig4} and the diagrams of \citeauthor{tin03} may exist for these types.}  
The curves in Figures~\ref{Golimowski.fig4}a and \ref{Golimowski.fig4}b are low-order polynomials fitted to the unweighted
data, excluding those for known close binaries.  The coefficients and residual statistics of these fits are listed
in Table~4.  The fits are not $\chi^2$-optimal, but they provide means of estimating the luminosities and distances
of M, L, and T dwarfs for which MKO $L'M'$ photometry and near-infrared spectral types have been obtained.  The 
combination of $L'M'$ photometry with $JHK$-based spectral types seems awkward, but the relatively small scatter in
the $L'$ and $M'$ luminosities -- especially among types L0 to L5 -- provides an advantage over combinations of 
$JHK$ photometry and optical or near-infrared spectral types (L02; D02; \citealt{tin03}; V04; K04).

The fits in Figures~\ref{Golimowski.fig4}a and \ref{Golimowski.fig4}b indicate that $M_{L'}$ and $M_{M'}$ decrease monotonically
throughout the M, L, and T classes.  They do not exhibit the pronounced ``hump'' or inflection in luminosity 
noted for early-T dwarfs in diagrams of $M_{I_C}$, $M_Z$, $M_J$, $M_H$, and $M_K$ versus spectral type 
(\citealt{tin03}; V04; K04).  The amplitude of this feature increases from $I_C$ to $J$ \citep{tin03} and decreases
from $J$ to $K$ (V04; K04).  The lack of obvious humps or inflections in Figures~\ref{Golimowski.fig4}a and 
\ref{Golimowski.fig4}b suggests that the latter trend continues through the $L'$ and $M'$ bands.  Close inspection of 
Figure~\ref{Golimowski.fig4}a shows that the polynomial fit may overestimate by $\sim 0.5$~mag the values of $M_{L'}$
for early-T dwarfs, but more $L'$ and parallax measurements of T0--T5 dwarfs are needed to confirm this 
possibility.  No inflection appears in Figure~\ref{Golimowski.fig4}b, but insufficient data exist for types L6--T5 to 
assert with confidence that $M_{M'}$ is strictly monotonic within this range of types.

The apparently monotonic decrease of the $L'$ and $M'$ luminosities with decreasing $T_{\rm eff}$ is consistent 
with recent explanations of the ``early-T hump'' at shorter wavelengths.  Using models of precipitating condensate 
clouds, \citet{ack01} and \citet{mar02} showed that a horizontally-uniform cloud deck forms progressively deeper 
in the atmosphere and becomes more optically thick as $T_{\rm eff}$ decreases.  This behavior significantly affects
the $z$- through $K$-band fluxes of late-L and early-T dwarfs ($T_{\rm eff} \approx 1450$~K; see \S4.3), but it 
affects much less the emergent flux outside these bandpasses.  The migration of the cloud deck into the convective 
region of the atmosphere may also disrupt the deck's uniformity, thereby allowing more $J$-band flux from hotter 
layers of the atmosphere to escape through holes in the clouds \citep{ack01,bur02b}.  Alternatively, the efficiency of 
sedimentation may rapidly increase at the L--T transition and enhance the $J$-band flux (K04).  \citet{tsu03} 
also attributed the L--T transition to the inward migration of thin dust clouds as $T_{\rm eff}$ decreases, but they 
viewed the reported brightening of the $J$-band flux as an artifact of a small sample of brown dwarfs with different
masses, ages, and cooling tracks.  They did not extend their demonstration to shorter
or longer wavelengths, but \cite{tsu02} reported that the effect of cloud migration on the emergent spectrum is 
largest in the $J$ band.  Whether or not cloud migration alone is sufficient to explain the sudden appearance of
the ``early-T hump'', the dynamics of the cloud deck below $T_{\rm eff} \approx 1400$~K have comparatively little 
impact on the emergent $L'$ and $M'$ fluxes.  The predicted effects of temperature, clouds, and gravity on 
$M_{L'}$ and $M_{M'}$ are examined further in \S5.

\subsection{Bolometric Luminosities and Effective Temperatures}

Our large and comprehensive set of $L'$ measurements permits us to determine with reasonable 
accuracy the bolometric luminosities ($L_{\rm bol}$) and $T_{\rm eff}$ of ultracool dwarfs.  We have 
computed and compiled $L_{\rm bol}$ of 42 dwarfs in our sample for which flux--calibrated spectra, $L'$ 
photometry, and trigonometric parallaxes are available.  To this group, we have added nine M, L, and T 
dwarfs for which spectra and $JHK$ photometry exist and for which trigonometric parallaxes have recently 
been measured.  The $L'$ luminosities of these supplemental dwarfs can be estimated from the measured $L'$
magnitudes of dwarfs in our sample that have similar spectral types and $JHK$ colors (K04).  The 
names, spectral types, parallaxes, and magnitudes of these nine dwarfs are listed in Table~5 along with 
their respective references.

As a first step toward computing $L_{\rm bol}$, we used SDSS $iz$, UKIRT $Z$, and MKO $JHK$ photometry 
(L02; K04) to calibrate the 0.8--$2.5~\mu$m spectra (\citealt{leg99}; G02; \citealt{bur02a}; K04) of the 
51 dwarfs under study.  We also used our MKO $L'$ and $M'$ measurements of Gl~229B to calibrate its 
3.0--$4.2~\mu$m and 4.5--$5.1~\mu$m spectra \citep{nol97,opp98}.  For Gl~229B, we summed the available 
spectra from $I$ through $M$ bands, linearly interpolated the fluxes in the regions 2.5--$3.0~\mu$m 
and 4.2--$4.5~\mu$m, and assumed a Rayleigh--Jeans (R--J) flux distribution longward of $M$.  The R--J
approximation is compromised by the presence of absorption by CH$_4$, H$_2$O, and NH$_3$ between 6 and 
$11~\mu$m \citep{mar96,bur01}, but we estimate that this absorption decreases by $\lesssim 1$\% the 
bolometric flux of dwarfs with $T_{\rm eff} \gtrsim 600$~K \citep{bur01}.  For the other dwarfs, 
we summed the spectra from their blue limits through $K$ band, interpolated the flux between $K$ 
and the effective $L'$ flux computed from our photometry, and assumed a R--J distribution longward of 
$L'$.  Neither the interpolation between $K$ and $L'$ nor the R--J extrapolation longward of $L'$ is 
valid for T dwarfs, because CH$_4$ absorbs shortward of $L'$ and CO absorbs significantly in $M$.
Consequently, we used the summed $L$- and $M$-band spectra of Gl~229B to determine corrections for 
these approximations.  The corrections {\it increase} by 20\% the derived bolometric fluxes of mid- 
to late-T dwarfs.  (The absorption by CH$_4$ and CO is more than offset by the flux beyond $L'$ that 
exceeds our R--J approximation.)  \citet{leg01} used model atmospheres to determine that no correction 
is needed for late-M to mid-L dwarfs.  For types L8--T3.5, we adopted a correction that is half that 
computed for the later T dwarfs.  We applied these corrections to the fluxes of the mid-L to late-T
dwarfs, verifying where possible the bolometric fluxes derived from this method against those computed 
using our $M'$ measurements and an R--J approximation longward of $M'$.  We found that the two methods 
matched within $\sim 5$\%.  We estimate that the uncertainties in the bolometric fluxes of all the 
dwarfs are 5--11\%.

We used the magnitudes obtained from our bolometric fluxes and the $K$ photometry of L02 and K04 to 
compute $K$-band bolometric corrections (BC$_K$) for the 51 M, L, and T dwarfs in our supplemented sample.
We also used the weighted-mean parallaxes listed in Tables~1 and 5 to convert the bolometric fluxes
into $L_{\rm bol}$ and compute absolute bolometric magnitudes ($M_{\rm bol}$).  Table~6 lists $L_{\rm bol}$ 
[expressed as log$(L_{\rm bol}/L_{\odot})$], $M_{\rm bol}$, and BC$_K$ for the 51 dwarfs.  Many of 
these quantities are based on photometry and astrometry reported since the work of L02 (D02; 
\citealt{tin03}; V04; K04; this paper), so the information in Table~6 supersedes that given in 
Table~7 of L02.

We used the relationships between $L_{\rm bol}$ and $T_{\rm eff}$ derived from the evolutionary models of 
\citet{bur97}, \citet{bar98}, and \citet{cha00} to obtain $T_{\rm eff}$ for the dwarfs listed in Table~6.
Because the radii of brown dwarfs older than 0.1~Gyr vary by no more than 30\% \citep{mar96,bur01}, the range
of possible $T_{\rm eff}$ for a given $L_{\rm bol}$ remains within $\sim 300$~K regardless of mass or age.
This behavior is illustrated in Figure~\ref{Golimowski.fig5}, which shows the relationships between $L_{\rm bol}$ 
and $T_{\rm eff}$ predicted by the models of \citet{bur97} and \citet{cha00}.  Column~6 of Table~6 lists the 
range of $T_{\rm eff}$ derived from the values of $L_{\rm bol}/L_{\odot}$ listed in Column~5, assuming 
ages of 0.1--10~Gyr.\footnote{The ages of some dwarfs have been further constrained by assuming coevality with
their main sequence companions whose ages have been delimited observationally.  These systems are: Gl~229AB 
(0.5--10~Gyr; \citealt{nak95}; J.~Stauffer 2001, personal communication; \citealt{leg02b,giz02}; I.~N.~Reid 
2003, personal communication), LHS~102AB \citep[1--10~Gyr;][]{leg02b,giz02}, GD~165B \citep[1.2--5.5~Gyr;][]{kir99a},
Gl~584C \citep[1--2.5~Gyr;][]{kir01}, and Gl~570D \citep[2--5~Gyr;][]{bur00,geb01}.  Also, the age of Kelu-1 
has been constrained to 0.3--1~Gyr on the basis of its Li~{\scriptsize I} $\lambda6708$~\AA\ absorption 
strength \citep{bas98}.}  Column~7 lists $T_{\rm eff}$ for an age of $\sim 3$~Gyr (unless otherwise noted), 
which represents the mean age of nearby brown dwarfs inferred from their kinematics (D02).  These values of
$T_{\rm eff}$ supersede those reported by L02 for dwarfs common to both samples.  The broad range of assumed
ages contributes uncertainties of $\sim 10$\% to $T_{\rm eff}$.   For dwarfs whose computed values of 
$L_{\rm bol}$ have errors within 10\%, the contributions of these errors to the uncertainties in $T_{\rm eff}$
are 1--2.5\%.  Thus, the uncertainties in $T_{\rm eff}$ for dwarfs whose measured parallaxes have errors 
$\lesssim 5$\% are dominated by our conservative range of ages for the dwarfs.  A less conservative range of 
0.5--10~Gyr increases the minimum $T_{\rm eff}$ for each dwarf by $\sim 200$~K.

Figures~\ref{Golimowski.fig6}a and \ref{Golimowski.fig6}b are diagrams of BC$_K$ and $T_{\rm eff}$ versus spectral
type for the M6--T9 dwarfs listed in Table~6.   The plotted values of $T_{\rm eff}$ are those listed in 
Column~7 of the table for a mean age of $\sim 3$~Gyr, except where noted.  The error bars for these values 
reflect the full ranges of $T_{\rm eff}$ listed in Column~6 of the table.  The curves in 
Figures~\ref{Golimowski.fig6}a and \ref{Golimowski.fig6}b are nonoptimal fourth- and sixth-order polynomials fitted 
to the weighted data, excluding the data for known close binaries.  The datum for the anomalous T6 dwarf 
2MASS~0937+2931 (see \S4.5) is omitted from in Figure~\ref{Golimowski.fig6}a for clarity's sake, but it is 
included in the polynomial fit to the data.  The fit in Figure~\ref{Golimowski.fig6}b is fixed at type T9 to 
avoid an unrealistic upturn in $T_{\rm eff}$ between types T8 and T9.  The coefficients and residual 
statistics of these fits are listed in Table~4.  

The fitted curve in Figure~\ref{Golimowski.fig6}a shows that BC$_K$ is a piecewise-monotonic function of spectral type 
with a small dispersion ($\sim 0.1$~mag) for types M6--T5.  The increased dispersion for the late-T dwarfs may 
indicate the sensitivity of H$_2$ CIA, which significantly affects the $K$-band spectrum, to variations in surface 
gravity (\citealt{sau94,bur02a}; K04).  BC$_K$ gradually rises between types M5 and L5, which is expected from Wien's 
law but may also indicate enhanced $K$-band luminosity as the cloud deck settles below the ``$K$-band photosphere'' 
\citep[$1500 \lesssim T_{\rm eff} \lesssim 1400$~K;][]{ack01,mar02}.  BC$_K$ generally declines for types later than 
L5, which reflects the increasing strength of combined absorption by CH$_4$ at 2.2--$2.4~\mu$m and CO at 
2.3--$2.5~\mu$m \citep[G02;][]{bur02a}.

Figure~\ref{Golimowski.fig6}b shows that $T_{\rm eff}$ declines steeply and monotonically for types M6--L7 and T4--T9.
The decline from L0 to L7 ($2300 \gtrsim T_{\rm eff} \gtrsim 1450$~K, for assumed ages of $\sim 3$~Gyr) is nearly 
linear, as noted by \citet{ste01}.  $T_{\rm eff}$ is approximately constant ($\sim 1450$~K) for types L7--T4, which
lie within the range of types for which $K$--$L'$ and $L'$--$M'$ appear constant (Figure~\ref{Golimowski.fig1}).  This 
coincidence suggests that the constancy of $T_{\rm eff}$ is the cause for these unchanging colors, but the substantial
changes in the $K$-band spectra of these brown dwarfs \citep[G02;][]{bur02a} show that their spectral energy 
distributions are not static at $\sim 1450$~K.  Indeed, the constancy of $T_{\rm eff}$ for types L7--T4 is not evident
in diagrams of $z$--$J$, $J$--$H$, and $H$--$K$ versus spectral type (L02; \citealt{bur02a,haw02}; K04).  These colors
increase or decrease substantially over this range.  The dichotomy between these changing colors and the nearly 
constant $K$--$L'$ and $L'$--$M'$ between types L6 and T5 may be attributed to the migration and disruption of 
condensate clouds deep in the photosphere \citep{ack01,mar02,bur02b,tsu02,tsu03}.  These cloud dynamics occur over a 
narrow range of $T_{\rm eff}$\footnote{\citet{bur02b} reproduced the 2MASS magnitudes and colors of L and T dwarfs by 
assuming that the cloud deck disrupts rapidly at $T_{\rm eff} \approx 1200$~K.  Likewise, K04 showed that the MKO $J$
and $K$ CMDs of L--T transition dwarfs are bounded by the 1300~K isotherms that connect the color--magnitude sequences 
predicted by cloudy and cloud-free models for a wide range of gravity.  These transition temperatures are $\sim 10$--15\%
lower than the $T_{\rm eff} \approx 1450$~K that we compute empirically for L7--T4 dwarfs using the effective temperatures
at age $\sim 3$~Gyr listed in Table~6.} and significantly affect only the 0.9--2.5~$\mu$m region of the flux spectrum.
Consequently, $L'$--$M'$ should not vary significantly across the L--T boundary.  Nevertheless, the small decrease in 
the $L'$ and $M'$ luminosities between types L6 and T5 (Figure~\ref{Golimowski.fig4}) suggests that some redistribution of 
spectral energy from 3.5--5.0~$\mu$m to shorter wavelengths occurs as the cloud deck settles or disrupts.  Thus, the 
constancy of $K$--$L'$ may be attributed to the balanced effects of enhanced $K$-band flux and increased CH$_4$ 
absorption at 2.2--$2.4~\mu$m, both of which occur rapidly as the cloud deck sinks, and gradually decreasing $L'$ 
luminosity.  

Figures~\ref{Golimowski.fig7}a and \ref{Golimowski.fig7}b show the variations of BC$_K$ with MKO $J$--$K$ and $K$--$L'$.  The 
data for 2MASS~0937+2931 are again omitted from the figures for clarity's sake.  These diagrams are augmented 
versions of ones shown by L02, and employ new $J$ photometry reported by K04.  BC$_K$ is neither a 
monotonic nor single-valued function of $J$--$K$ because of the color reversal brought on by increasing CH$_4$
absorption at 2.2--$2.4~\mu$m for types L8 and later (G02).  BC$_K$ is a better behaved function of $K$--$L'$, but 
it is degenerate for $K$--$L' \approx 1.6$.  This degeneracy reflects the balanced effects of flux redistribution 
and CH$_4$ absorption in the $K$- and $L'$-bands for L6--T5 dwarfs.  Thus, Figures~\ref{Golimowski.fig7}a and 
\ref{Golimowski.fig7}b are not useful stand-alone references for bolometric luminosities near the L--T boundary.

\subsection{Comparison of Effective-Temperature Scales}

Since the initial discoveries of numerous L dwarfs by DENIS and 2MASS, many estimates of the relationship
between $T_{\rm eff}$ and L subtype have been reported.  \citet{kir99b} and \citet{rei99} offered initial
estimates of the $T_{\rm eff}$ scale of L dwarfs by comparing the evolutions of absorption features in their 
optical spectra with chemical-equilibrium abundance profiles predicted for the atoms and molecules responsible
for those features \citep{bur99}.  \citet{mar99b} and \citet{bas00} derived a warmer $T_{\rm eff}$ scale by 
fitting synthetic absorption profiles of Rb~{\small I} $\lambda7948$~\AA\ and Cs~{\small I} $\lambda8521$~\AA\
with those observed in their optical spectra.  \citet{nol00}, \citet{leg01}, and \citet{sch02} also developed 
$T_{\rm eff}$ scales for L dwarfs by fitting model spectra to their sets of optical and near-infrared spectra.  
Consequently, the accuracy of each $T_{\rm eff}$ scale is tied to the fidelity of the contemporaneous model 
atmospheres on which the scale is based.  Differences among these scales probably reflect the rapidly evolving 
details of the model atmospheres rather than fundamentally different perspectives on the effective temperatures
of ultracool dwarfs.  Fortunately, empirically-based $T_{\rm eff}$ scales have been derived that are immune to
the idiosyncracies of model atmospheres and depend only on the comparatively robust theoretical relationship 
between the ages and radii of brown dwarfs (\citealt{leg01}; L02; D02; V04; this paper).  We now
compare the effective temperatures listed in Table~6, which supersede the results of \citet{leg01} and L02, 
with the empirical $T_{\rm eff}$ scales derived by D02 and V04.

D02 derived $T_{\rm eff}$ for 17 M and L dwarfs listed in Table~6.\footnote{The 17 dwarfs are LHS~3003 (M7), 
LHS~2065 (M9), BRI~0021--0214 (M9.5), 2MASS~J0345+2540 (L1), 2MASS~J1439+1929 (L1), 2MASS~J0746+2000AB (L1 + $\sim$L2),
DENIS~J1058--1548 (L3), GD~165B (L3), Kelu-1 (L3), 2MASS~J2224--0158 (L3.5), 2MASS~J0036+1821 (L4), LHS~102B (L4.5),
2MASS~J1507--1627 (L5.5), DENIS~J0205--1159AB (L5.5 + L5.5), 2MASS~J0825+2115 (L6), DENIS~J1228--1547AB (L6 + $\sim$L6),
and 2MASS~J1632+1904 (L7.5).}  In doing so, they applied the BC$_{K_{\rm UKIRT}}$ versus $I_{\rm C}$--$K_{\rm UKIRT}$
relation of \citet{leg01} to their collection of $JHK$ photometry measured on the California Institute of Technology
(CIT) and 2MASS photometric systems.  They also adopted radii that are halfway between those predicted by the models 
of \citet{bur97} and \citet{cha00} for their resultant values of $M_{\rm bol}$ and assumed ages of 1--5~Gyr.  The 
values of $T_{\rm eff}$ derived by D02 for the eleven M7--L4.5 dwarfs are higher by an average of $\sim 60$~K than 
those listed in Column~7 of Table~6 for the same dwarfs.  Conversely, the values derived by D02 for the five L5.5--L7.5 
dwarfs are lower by an average of $\sim 30$~K than the corresponding values in Table~6.  Although these discrepancies 
lie within the ranges of uncertainty of both $T_{\rm eff}$ scales, their systematic nature is likely due to slightly
different applications of the evolutionary models for particular combinations of luminosity and age.  We discount 
the possibilities that these differences are caused by discordant photometric measurements or improper use of the 
UKIRT bolometric corrections by D02.

The ranges of $T_{\rm eff}$ computed by D02 for the 17 dwarfs are approximately half as wide as those listed in Column~6
of Table~6 for the same dwarfs.  D02's smaller uncertainties are not the result of more accurate data, but instead reflect
the 1--5~Gyr range of ages assumed for all the dwarfs in their sample.  This range is much narrower than our adopted range
of 0.1--10~Gyr for the dwarfs listed in Table~6 whose ages cannot be constrained by the ages of stellar companions.  By 
assuming a lower age limit of 1~Gyr, D02 eliminate from consideration the era in which the radii of ultracool dwarfs change
greatly and rapidly \citep{bur97, cha00}.  The possible youth of Kelu-1 and Gl~229B$^{13}$ suggest that the 1--5~Gyr range
of ages assumed by D02 is too narrow to encompass a random sample of ultracool dwarfs in the solar neighborhood.  In general,
the ranges of ages assumed for such samples must be carefully considered when comparing $T_{\rm eff}$ scales derived from 
structural models.  Figure~\ref{Golimowski.fig5} shows that, for a fixed $L_{\rm bol}$, narrowing the age range from 0.1--10~Gyr 
to 1--5~Gyr compresses the corresponding range of $T_{\rm eff}$ asymmetrically so that its midpoint shifts to a higher
$T_{\rm eff}$ than would be expected if the radii of brown dwarfs decreased uniformly over time.  Consequently, comparisons
of $T_{\rm eff}$ scales must be based on temperatures derived for some fiducial age or radius, rather than the midpoint of 
the $T_{\rm eff}$ range.  Otherwise, discrepancies between $T_{\rm eff}$ scales might be declared where none actually exists.  

V04 applied our polynomial fit of BC$_K$ versus spectral type (Table~4) to a sample of 56 L and T dwarfs whose trigonometric 
parallaxes have been measured at the United States Naval Observatory (D02; V04).  In doing so, they assumed equality between
the dwarfs' $K$ magnitudes, which were collected from different sources and transformed to approximate $K_{\rm CIT}$ magnitudes,
and the $K_{\rm MKO}$ magnitudes on which our bolometric corrections are based.  V04 also employed spectral types based on the
optical L-dwarf classification scheme of \citet{kir99b} and the near-infrared T-dwarf classification scheme of \citet{bur02a},
rather than the near-infrared classification scheme of G02 that defines our L and T subtypes.  Despite the noted differences 
between the CIT and MKO photometric systems \citep{ste04} and the optical and near-infrared classification schemes \citep{ste03},
the values of log$(L_{\rm bol}/L_{\odot})$, $M_{\rm bol}$, and $T_{\rm eff}$ obtained by V04 are generally consistent with those
shown in Table~6 for the dwarfs common to both studies.  However, significant differences exist for some individual dwarfs and
spectral types.  For instance, our ranges of $T_{\rm eff}$ for early-L dwarfs are 100--400~K cooler than those of V04.  This 
discrepancy is caused by the fixed range of radii (0.075--0.105~$R_{\odot}$) adopted by V04 for all the dwarfs in their sample.
Imposing a less conservative, but more appropriate, lower limit of $\sim 0.1~R_{\odot}$ upon dwarfs with 
log$(L_{\rm bol}/L_{\odot}) \gtrsim -4.5$ \citep{bur97,cha00} brings V04's $T_{\rm eff}$ scale for early-L dwarfs into agreement
with ours.  V04 computed log$(L_{\rm bol}/L_{\odot}) = -5.58 \pm 0.10$ and $T_{\rm eff} = 764^{+88}_{-71}$~K for the T9 dwarf 
2MASS~J0415--0935.  These values are significantly higher than the corresponding values in Table~6.  The discrepancies are 
due to a 0.4-mag difference between our measured $K_{\rm MKO}$ magnitude and the transformed $K_{\rm CIT}$ magnitude 
adopted by V04.  The 0.4-mag difference is probably caused by the 20\% uncertainty in the measured 2MASS $K_s$ magnitude of 
2MASS~J0415--0935 and the \hbox{$\sim 0.2$-mag} systematic error for late-T dwarfs associated with the 2MASS-to-CIT transformation
employed by V04 \citep{ste04}.  Nevertheless, the possibility that 2MASS~J0415--0935 is photometrically variable cannot be excluded.

\subsection{Noteworthy L and T dwarfs}

Several ultracool dwarfs in our sample merit special consideration.  We describe them here, in progressive
order of spectral type.

{\it Kelu-1 (L3)} is $\sim 1$~mag more luminous in $L'$ and $M'$ than the other L3 dwarfs in our sample.  
Similar overluminosity in other bandpasses is well documented \citep{mar99b,leg01}.  Kelu-1's large rotational 
velocity \citep[$60 \pm 5$~km~s$^{-1}$;][]{bas00} and periodic photometric variability \citep{cla02} suggest 
possible duplicity, but no companion has yet been imaged \citep{mar99a}.  Kelu-1's age has been constrained 
to 0.3--1~Gyr based on the strength of Li~{\small I} $\lambda6708$~\AA\ absorption \citep{bas98}.  For this range 
of ages, our computed $L_{\rm bol}$ yields $T_{\rm eff} = 2100$--2350~K.  These temperatures are $\sim 400$~K hotter
than those the other L3 dwarfs.  For Kelu-1 to have a $T_{\rm eff}$ consistent with the other L3 dwarfs, it must 
have a mass of $\sim 0.012$~$M_{\odot}$ and an age of $\sim 10$~Myr \citep{bur97}.  This age is inconsistent
with the lower bound set by the Li~{\small I} absorption, and Kelu-1 is not located near a known region of star 
formation \citep{rui97}.  Moreover, Kelu-1 exhibits H$\alpha$ emission, which is characteristic of old, early-L 
dwarfs \citep{giz00}.  These conditions don't preclude the possibility that Kelu-1 is extremely young, 
but the collective evidence favors unresolved duplicity as the cause of Kelu-1's overluminosity.

{\it 2MASS~J2244+2043 (L7.5)} has a $K$--$L'$ color that is $\sim 0.3$~mag redder than those of other late-L dwarfs.
D02 reported that its 2MASS $J$--$K_s$ color is $\gtrsim 0.5$~mag redder than those of all other L dwarfs 
in their sample.  K04 reported that the MKO $J$--$H$ and $H$--$K$ colors of 2MASS~J2244+2043 are significantly 
redder than those of other L dwarfs whose spectral types lie within the broad range (L5.5--L9.5) spanned by the
near-infrared spectral indices (G02) computed for 2MASS~J2244+2043.  K04 suggested that the anomalous $JHK$ colors 
may be caused by condensate clouds that are more optically thick than usual.  A comprehensive set of $J$- through 
$L$-band spectra of 2MASS~J2244+2043 is needed to determine whether unusually opaque clouds or other conditions 
cause the excessively red colors throughout these bands.

{\it SDSS~J0805+4812 (L9)} is an anomalously blue ($K$--$L' = 1.20$) late-L dwarf.  K04 reported that its $J$--$H$ 
and $H$--$K$ colors are $\sim 0.2$--0.3~mag bluer than those of other dwarfs whose spectral types lie within the
L7.5--T0.5 range spanned by the indices (G02) computed for SDSS~J0805+4812.  Its 1.0--$2.5~\mu$m spectrum reveals 
unusually strong H$_2$O, K~{\small I}, and FeH absorption, which suggests that the atmosphere of SDSS~J0805+4812 
is relatively free of condensate clouds or metal-poor (K04).  The former possibility is inconsistent with the 
observed value of $K$--$L'$, however, because a cloudless atmosphere with $T_{\rm eff} \approx 1400$~K should 
yield a redder value of $K$--$L'$ than a corresponding cloudy atmosphere (see \S5).  A comprehensive set of $J$- 
through $L$-band spectra is needed to determine the cause(s) of the unusual colors of SDSS~J0805+4812.

{\it SDSS~J0423--0414 (T0)} is $\sim 1$~mag more luminous in $L'$ and $M'$ than other dwarfs of similar spectral type 
in our sample.  Overluminosities of $\sim 0.75$--1.5~mag in $J$, $H$, and $K$ have also been reported by V04 
and K04.  SDSS~J0423--0414 is not known to be multiple.  V04 state that its $JHK$ colors and luminosities 
better suit its optical spectral type of L7.5 \citep{cru03} than its near-infrared spectral type of T0 (G02).  This 
contradiction may be virtual, however, because the spectral classification schemes of \citet{kir99b} and G02 are not 
rigidly correlated for late-L and early-T dwarfs.  Unfortunately, $K$--$L'$ and $L'$--$M'$ are nearly constant for 
near-infrared types L7--T4 (Figure~\ref{Golimowski.fig1}), so these colors do not constrain SDSS~J0423--0414's spectral type.
However, Figures~\ref{Golimowski.fig4}a and \ref{Golimowski.fig4}b show that the $L'$ and $M'$ overluminosities of this dwarf are the same
whether it has a near-infrared type of L7.5 or T0.  Moreover, Table~6 and Figure~\ref{Golimowski.fig6}a show that the BC$_K$
computed for SDSS~J0423--0414 is more consistent with type T0 than type L7.5.  Thus, our photometry and derived BC$_K$ 
support the T0 classification assigned to SDSS~J0423--0414 by G02 on the basis of its highly consistent near-infrared 
spectral indices.  Our results do not refute the L7.5 optical classification; they merely reflect the dominant
contribution of the dwarf's near-infrared flux to its bolometric flux.  Our computed $L_{\rm bol}$ for SDSS~J0423--0414 
yields $T_{\rm eff} = 1450$--1825~K for assumed ages of 0.1--10~Gyr.  These temperatures are $\sim 300$~K hotter than 
those of other dwarfs with types L9--T1 (Table~6).  For SDSS~J0423--0414 to have a $T_{\rm eff}$ 
consistent with the other L9--T1 dwarfs, it must have a mass of $\sim 0.009$~$M_{\odot}$ and an age of $\sim 3$~Myr 
\citep{bur97}.  \citet{bur03a} speculate that SDSS~J0423--0414 is an older, more massive T dwarf because it exhibits 
H$\alpha$ emission and it has an optical continuum whose slope is consistent with relatively large surface gravity.  
Moreover, SDSS~J0423--0414 does not lie near a known star-forming region.  Thus, the collective evidence indicates that 
SDSS~J0423--0414's overluminosity is likely caused by unresolved multiplicity rather than extreme youth.  Its putative 
components probably have equal masses, because a coeval companion of lesser mass would have $T_{\rm eff} \lesssim 
1450$~K and a spectral type later than T4 (Figure~\ref{Golimowski.fig6}b).  SDSS~J0423--0414's near-infrared spectrum 
does not exhibit such heterogeneity (G02). 

{\it 2MASS~J0559--1404 (T4.5)} was reported by D02 to be $\sim 1$~mag more luminous in $J$
than the L8 dwarfs Gl~337C, Gl~584C, and 2MASS~J1632+1904.  {\it Hubble Space Telescope} observations revealed
no bright companion beyond 0\farcs05 of the T dwarf \citep{bur03c}.  Our $L'$ and $M'$ measurements
indicate that 2MASS~J0559--1404 is no more luminous than the upper bound of the intrinsic scatter observed
for the T dwarfs.  This conclusion agrees with that of V04, who find that the $J$, $H$, and
$K$ luminosities of 2MASS~J0559--1404 are consistent with the ``early-T hump.''

{\it 2MASS~J0937+2931 (T6)} has an anomalously red color of $K$--$L' = 3.05$.  Its $M_{L'}$ is consistent 
with those of other T6 dwarfs (Table~3), so its overly red $K$--$L'$ can be attributed to its suppressed 
$K$-band flux caused by uncommonly strong H$_2$ CIA \citep{bur02a}.  Such strong absorption may be 
attributed to high surface gravity (log~$g > 5.5$; K04) or to low metallicity (\citealt{sau94,bur02a}; K04).
Either condition may also account for the absence of the K~{\small I} absorption doublet at $1.24~\mu$m 
and $1.25~\mu$m in 2MASS~J0937+2931's $J$-band spectrum (\citealt{bur02a}; K04), because low metallicity 
implies a paucity of sodium and because high gravity raises the abundance of KCl at the expense of K 
(\citealt{lod99,mar02}; K04).  Recent models of the pressure-broadened Na~{\small I} and K~{\small I} 
absorption lines in the 0.6--1~$\mu$m spectrum of 2MASS~J0937+2931 suggest that a mixed condition of high 
gravity and low metallicity best describe this unusual T dwarf \citep{bur03a}.

{\it 2MASS~J0415--0935 (T9)} is the latest T dwarf classified on the system of G02 (K04).  It has
been classified as type T8 by \citet{bur02a}, but its numerical rank based on the average of its spectral 
indices is the latest of the T dwarfs in their sample.  Our computed values of log$(L_{\rm bol}/L_{\odot})
= -5.73 \pm 0.05$ and $T_{\rm eff} = 600$--750~K show that 2MASS~J0415--0935 is the least luminous and
coolest brown dwarf presently known.  It is 35--225~K cooler than the previous title-holder, Gl~570D 
\citep{bur00,geb01}.  Interestingly, the $J$--$H$ and $H$--$K$ colors of 2MASS~J0415--0935 are redder than 
those of Gl~570D (K04), which is contrary to the trend that later T dwarfs have bluer colors in these bands.
\citet{mar02} and \citet{bur03} predict a reversal in $J$--$K$ as H$_2$O condenses and settles into clouds 
at $T_{\rm eff} \lesssim 500$~K.  If the redder $JHK$ colors of 2MASS~J0415--0935 are caused by thickening 
water clouds, then our computed value of $T_{\rm eff}$ indicates that the condensation of H$_2$O can occur 
under warmer conditions than anticipated from typical model atmospheres.  However, \citet{bur03} show that
H$_2$O clouds can form at such temperatures in the atmospheres of very old ($\sim 6$--10~Gyr) brown dwarfs with 
masses $\sim 0.040$--$0.060~M_{\odot}$.  Alternatively, the color reversal may be due to optically-thick 
clouds that conceivably form when gaseous potassium condenses into solid KCl at $T_{\rm eff} \approx 600$~K 
\citep{lod99,mar00,bur03}.  Further study of 2MASS~J0415--0935 is needed to investigate these possibilities.

\section{Effects of Precipitating Clouds, Nonequilibrium Chemistry, and Gravity}

Our theoretical understanding of the effects of condensate clouds on the emergent fluxes of brown dwarfs has 
advanced considerably during the last few years \citep{bur97,cha00,all01,ack01,mar02,bur02b,tsu02, tsu03,
coo03,bur03}.  The formation, migration, sedimentation, and turbulent disruption of cloud decks are thought 
to affect significantly the near-infrared spectra of L and T dwarfs, as well as cooler dwarfs yet to be 
discovered.  The models of \citet{ack01} and \citet{mar02} consider horizontally-uniform decks of 
precipitating water, iron, and silicate clouds formed in atmospheres having solar metallicity and conditions
of structural and thermochemical equilibrium.  The altitudes, particle-size distributions, and density profiles
of the clouds are determined self-consistently from atmospheric temperature and pressure profiles and an 
adjustable ratio, $f_{\rm sed}$, which describes the efficiency of particle sedimentation (precipitation) 
relative to the upward transport of condensates by convection.\footnote{\citet{ack01} originally employed 
the parameter $f_{\rm rain}$ to describe the sedimentation efficiency.  To avoid confusion with the traditional
notion of rain, $f_{\rm rain}$ has been renamed $f_{\rm sed}$.}  Practical values of $f_{\rm sed}$ for L dwarfs
range from 3, which also describes Jupiter's thick NH$_3$ cloud deck, to 5, which describes a thinner, more 
efficiently precipitating cloud deck (\citealt{ack01,mar02,bur02b}; K04).  Model spectra from $0.5~\mu$m to 
$5.0~\mu$m have been produced by \citeauthor{mar02}, and our $L'$ and $M'$ measurements allow us to assess 
these models between $3~\mu$m and $5~\mu$m. 

Figures~\ref{Golimowski.fig8}, \ref{Golimowski.fig9}, and \ref{Golimowski.fig10} are diagrams of $M_K$, $M_{L'}$, and $M_{M'}$ versus 
$T_{\rm eff}$ for the L3--T9 dwarfs listed in Table~6.  The dwarfs for which we have $M'$ data are denoted by 
filled symbols to facilitate comparison of the data associated with these dwarfs in all three diagrams.  The 
absolute magnitudes of close binaries have been increased by 0.75~mag to represent one component of the presumed 
uneclipsed, equal-luminosity systems.  The curves in the diagrams are the predicted absolute magnitudes computed 
from the models of \citet{mar02} for discrete values of $f_{\rm sed}$ (3, 5, and no clouds) and surface gravity 
(log~$g = 4.5$, 5.0, and 5.5, where $g$ has units of cm s$^{-2}$).  The precipitating-cloud models are shown 
for $2000 \geq T_{\rm eff} \geq 1300$~K, and the cloud-free models are shown for $T_{\rm eff} \leq 1500$~K.  
Thus, all models are shown across the L--T transition.  The predicted magnitudes were synthesized from the model 
spectra and the measured transmission profiles of the MKO $K$, $L'$, and $M'$ filters (L02).  The diagrams show 
that the models collectively envelope the empirical data in all bands, i.e., the models reproduce the broad 
ranges of absolute magnitudes and effective temperatures for this sample of L and T dwarfs.  The ensemble of 
data does not favor particular sets of model parameters for particular ranges of spectral type, but this 
situation is expected given the presumably heterogeneous masses, ages, and metallicities associated with our 
sample.  

Figures~\ref{Golimowski.fig8}--\ref{Golimowski.fig10} indicate that the high-gravity (log~$g = 5.0$--5.5) models consistently 
match the absolute magnitudes and effective temperatures obtained for the L3--T4.5 dwarfs in our sample.  
Conversely, the values of $M_K$ and $M_{L'}$ for the T6--T9 dwarfs (except the anomalous 2MASS~J0937+2931)
are generally bounded by the low-gravity (log~$g = 4.5$--5.0) models.  The lower gravities of the late-T dwarfs
are also indicated in K04's Figure~6, which compares the measured MKO $J$--$H$ and $H$--$K$ colors of 42 T 
dwarfs with those synthesized from the models of \citet{mar02}.  The apparent fidelity of the low-gravity 
models suggests that they are useful benchmarks for predicting other photometric characteristics of late-T 
dwarfs.  Such reasoning has frequently been applied when estimating the fluxes of cool brown dwarfs at 
wavelengths longer than $4~\mu$m, which are easily observed from space.  Figure~\ref{Golimowski.fig10}, however, 
shows that the values of $M_M'$ measured for {\it all} the T dwarfs in our sample are better matched by the 
high-gravity models.  The log~$g = 4.5$--5.0 models, which consistently reproduce $M_K$ and $M_L'$ measured 
for the late-T dwarfs Gl~229B and 2MASS~J0415--0935, underestimate $M_{M'}$ for these dwarfs by 0.5--1~mag.
L02 also noted discrepancies of $\gtrsim 1$~mag between the measured $K$--$M'$ colors of two T dwarfs 
(SDSS~J1254--0122 and 2MASS~J0559--1404) and the $K$--$M'$ colors predicted by the dusty-atmosphere models 
of \citet{cha00} and the settled-condensate models of \citet{bur97} for wide ranges of gravity.  Noting that
these models predicted $M_K$ well for their sample of L and T dwarfs, L02 attributed the discrepancies to 
$M'$ luminosities that were overpredicted by a factor of $\sim 3$.  This assessment is consistent with the 
general trends seen in Figures~\ref{Golimowski.fig8}--\ref{Golimowski.fig10}, but the figures also show that 2MASS~J0559--1404's
$M_K$, $M_{L'}$, and $M_{M'}$ are consistent with the cloud-free, log~$g = 5.5$ model of \citet{mar02}.

Although the discrepancies between the observed and predicted $M'$ luminosities of most T dwarfs vary among the 
models, they reveal a consistent overestimation of the emergent $5~\mu$m flux regardless of how the condensates 
are modeled. L02 speculated that the low $M'$ luminosities of SDSS~J1254--0122 and 2MASS~J0559--1404 are caused 
by strong CO absorption at 4.5--$4.9~\mu$m, which was predicted and then observed in the $M$-band spectrum of 
Gl~229B \citep{feg96,nol97,opp98}.  The models of \citet{bur97}, \citet{cha00}, and \citet{mar02} do not account 
for this absorption because it results from an abundance of CO that exceeds that expected under conditions of 
thermochemical equilibrium.  \citet{sau03} have modeled the effects of nonequilibrium chemistry caused by vertical
mixing on the emergent spectrum of brown dwarfs.  They determined that the overabundance of CO in cloudless 
atmospheres significantly decreases the $M'$ fluxes from their chemical-equilibrium levels for $T_{\rm eff} 
\lesssim 1400$~K.  As Figure~\ref{Golimowski.fig6}b shows, this range of $T_{\rm eff}$ spans the late half of
the presently defined T sequence.

\citet{sau03} showed that the measured values of $M_{M'}$ for 2MASS~J0559--1404 and Gl~229B are matched by a 
nonequilibrium model with log~$g = 5$ and an eddy-diffusion coefficient of $\sim 100$~cm$^2$~s$^{-1}$.  Such a 
coefficient is consistent with the minimum expected for planetary atmospheres and indicates that the vertical 
mixing of CO occurs within the outermost radiative layer of the atmosphere.   An independent measure of 
2MASS~J0559--1404's gravity is needed, however, to resolve the ambiguity between the log~$g = 5$, nonequilibrium
model of \citet{sau03} and the cloud-free, log~$g = 5.5$, equilibrium model of \citet{mar02} shown in 
Figure~\ref{Golimowski.fig10}.  Our results for the T9 dwarf 2MASS~J0415--0935 provide a less ambiguous test of the 
nonequilibrium models for the coolest known T dwarfs ($T_{\rm eff} \approx 700$~K).  Figures~\ref{Golimowski.fig8} and 
\ref{Golimowski.fig10} show that the cloud-free, log~$g = 4.5$ model of \citet{mar02} matches $M_K$ and $M_{L'}$ well, 
but it underestimates $M_{M'}$ by $\sim 1$~mag.  Conversely, the log~$g = 5.5$ model predicts $M_{M'}$ within 
0.25~mag, but overestimates $M_K$ and $M_{L'}$ by 1.6~mag and 0.6~mag, respectively.  Figure~4 of \citet{sau03}
shows that the measured $M_{M'} = 14.03 \pm 0.15$ for 2MASS~J0415--0935 is well matched by a nonequilibrium, 
log~$g = 5$ model with a large eddy-diffusion coefficient ($\sim 10^4$~cm$^2$~s$^{-1}$) typical of planetary 
atmospheres.

Evolutionary models of ultracool dwarfs \citep{bur97,cha00} can be used to constrain the gravities and masses
of the dwarfs in our sample, if the dwarfs' ages can be estimated.  D02 used kinematic statistics to argue that
the mean age of the L and T dwarfs in the solar neighborhood in 2--4~Gyr.  We find an age range of 0.3--5.5~Gyr 
for the dwarfs in our sample whose minimum and maximum ages can be constrained spectroscopically.$^{13}$  The 
lower limit of this range is consistent with a reported lull in star formation in the solar neighborhood 
during the last $\sim 0.5$~Gyr \citep{her00,giz02}.  The evolutionary tracks of \citet{bur97} show that L2--T4 
dwarfs ($T_{\rm eff} \approx 2000$--1400~K) with ages 0.3--5.5~Gyr have log~$g \approx 5.0$--5.5.  This prediction
is consistent with the results shown in Figures~\ref{Golimowski.fig8}--\ref{Golimowski.fig10}.  The evolutionary tracks also show 
that T6--T9 dwarfs ($T_{\rm eff} \approx 1000$--700~K) in this age range have log~$g \approx 4.5$--5.3, but their
mean gravity is log~$g \gtrsim 5.0$ because high-mass brown dwarfs cool much more slowly than low-mass brown 
dwarfs \citep{rei99}.  This range of gravity is higher than the log~$g \approx 4.5$--5.0 noted for T6--T9 dwarfs 
in Figures~\ref{Golimowski.fig8} and \ref{Golimowski.fig9}.  Unfortunately, the small and heterogeneous nature of our sample precludes
a definitive explanation of this discrepancy.  Because the magnitudes of the late-T dwarfs are typically near the 
detection limits of 2MASS and SDSS, our sample may be biased toward younger, brighter, and less-massive ones.  
Alternatively, the gravities indicated by the solar-metallicity, equilibrium models of \citet{mar02} may be 
incorrect.  Nevertheless, the latter possibility does not affect our conclusions regarding the overpredicted $M'$
luminosities for late-T dwarfs.

\section{Implications for Spaced-Based Missions}

\citeauthor{bur97}\ (1997, 2001, 2003) created 1--30~$\mu$m spectra of brown dwarfs and extrasolar giant planets 
(EGPs) of various masses and ages using model atmospheres that assume settled condensate clouds and thermochemical 
equilibrium.  They found that the suppression of mid-infrared flux by H$_2$ enhances enormously the flux at shorter 
wavelengths.  For example, the $5~\mu$m flux of a 1~Gyr-old, Jupiter-mass EGP is $10^4$ times greater than its 
$T_{\rm eff} \approx 160$~K blackbody equivalent.  \citet{mar96} referred to this enhanced $5~\mu$m flux as the
``universal diagnostic'' of brown dwarfs and EGPs.  \citet{bur01} remarked that space-based, $M$-band imagers could
detect brown dwarfs cooler than can be found by DENIS, 2MASS, and SDSS.  \citet{bur03} added that the persistent 
$M$-band hump in the spectra of older and less massive brown dwarfs and EGPs makes this bandpass the best suited 
for studying these objects with the {\it Spitzer Space Telescope} ({\it SST}; formerly the {\it Space Infrared 
Telescope Facility}, or {\it SIRTF}).  Such searches for ``infra-T'' dwarfs and EGPs are indeed imminent now that 
{\it SST} has been launched (\citealt{pad03}; G.~Fazio 2003, personal communication\footnote{Presently, unpublished 
abstracts of the approved {\it SST} Guaranteed Time Observer science programs may be viewed on the World Wide Web at 
http://sirtf.caltech.edu/SSC/geninfo/gto/abs/.}).

The apparent 50--200\% overestimates of the $M$-band fluxes of T dwarfs by chemical-equilibrium models diminish
the anticipated sensitivity of the $4.5~\mu$m band of {\it SST's} Infrared Array Camera (IRAC) to the coolest known 
T dwarfs.  If the especially low $M'$ luminosity of 2MASS~J0415--0935 is indicative of low-mass brown dwarfs with
$T_{\rm eff} \lesssim 600$~K, then IRAC's $4.5~\mu$m detection horizons for nearby infra-T dwarfs and EGPs may be
significantly nearer than expected.  Moreover, the low-mass limits for members of young star clusters detected at
$4.5~\mu$m may be higher than anticipated.  The 1.0~$\mu$m width of IRAC's $4.5~\mu$m bandpass will mitigate 
somewhat the effect of the 4.5--$4.9~\mu$m CO absorption on the integrated signal-to-noise ratio, but it will 
also complicate the interpretation of the CO aborption strength.  By expanding our narrower-band $M'$ study to 
include more faint and cool brown dwarfs, we may aid the {\it SST} studies by calibrating the effects of CO 
absorption on the broader $4.5~\mu$m photometry of at least the warmer IRAC targets.

\citet{sau03} reported that nonequilibrium chemistry also affects the abundances of N$_2$ and NH$_3$ in the
atmospheres of cool brown dwarfs.  The observable effect of this situation is diminished absorption by NH$_3$
at $10.35~\mu$m and $10.75~\mu$m.  Thus, contrary to the case of CO in the $M$-band, vertical mixing serves to
enhance the $N$-band ($\sim 10~\mu$m) flux of brown dwarfs with $T_{\rm eff} \lesssim 1200$~K.  Unfortunately,
the $N$-band lies between the reddest bandpass of IRAC and the bluest bandpass of the Multiband Imaging 
Photometer for {\it SST} (MIPS), but targeted studies of extremely cool brown dwarfs with {\it SST's} Infrared 
Spectrograph (IRS) should benefit from their larger-than-predicted $10~\mu$m luminosities.  Moreover, future 
mid-infrared space missions like the {\it James Webb Space Telescope (JWST)} may fully exploit the enhanced 
$10~\mu$m luminosities of infra-T dwarfs and EGPs.  Despite the previously underappreciated effects of 
nonequilibrium chemistry in substellar atmospheres, the prospects for filling the ever-shrinking gap between 
the coolest known T dwarfs and the Jovian planets with {\it SST} and {\it JWST} are excitingly good.

\section{Summary}

Our compilation of new and previously reported MKO $L'$ and $M'$ photometry has permitted us to characterize
ultracool dwarfs comprehensively at wavelengths longward of the commonly used $J$, $H$, and $K$ bands.  We find 
that $K$--$L'$ increases monotonically with decreasing $T_{\rm eff}$, but the nearly constant $T_{\rm eff} \approx
1450$~K of spectral types L7--T4 limits the utility of $K$--$L'$ as an indicator of spectral type.  Likewise, 
$L'$--$M'$ is nearly constant between types L6 and T3, indicating that the dramatic changes in the 1--$2.5~\mu$m 
spectra of L--T transition dwarfs are not duplicated in their $L'$- and $M'$-band spectra.  This dichotomous 
behavior supports recent theories that the rapid migration, disruption, and/or thinning of condensate clouds at 
$T_{\rm eff} \lesssim 1400$~K occur at altitudes that are coincident with the regions of $z$- through $K$-band 
emission but are well below the $L'$-band and $M'$-band ``photospheres.''  The $L'$ and $M'$ luminosities of the 
early-T dwarfs do not exhibit the pronounced humps or inflections noted by others in the $I$ through $K$ bands, 
but insufficient data exist for types L6--T5 to assert that $M_{L'}$ and $M_{M'}$ are strictly monotonic within 
this range of types.

We used our $L'$ photometry, flux-calibrated $JHK$ spectra, and recently published trigonometric parallaxes
to compute $L_{\rm bol}$, BC$_K$, and $T_{\rm eff}$ for ultracool dwarfs.  We find that BC$_K$ is a well behaved 
function of spectral type with a dispersion of $\sim 0.1$~mag for types M6--T5.  BC$_K$ is significantly more 
scattered among the later T dwarfs, which may indicate the sensitivity of H$_2$ CIA in the $K$-band to varying 
surface gravity for $T_{\rm eff} \lesssim 1400$~K.  BC$_K$ is neither a monotonic nor single-valued function of 
$J$--$K$ because of the color reversal induced by the onset of CH$_4$ aborption at 2.2--$2.4~\mu$m at spectral type 
L8.  BC$_K$ is a single-valued function of $K$--$L'$ except at $K$--$L' \approx 1.6$, which corresponds to the 
L--T transition.  $T_{\rm eff}$ declines steeply and monotonically for types M6--L7 and T4--T9, but is nearly 
constant at $\sim 1450$~K for types L7--T4 with assumed ages of $\sim 3$~Gyr.  Our photometry and bolometric 
calculations indicate that Kelu-1 (L3) and SDSS~J0423--0414 (T0) are probable binary systems.  We compute 
log$(L_{\rm bol}/L_{\odot}) = -5.73 \pm 0.05$ and $T_{\rm eff} = 600$--750~K for 2MASS~J0415--0935 (T9), 
making it the least luminous and coolest brown dwarf presently known.

We have compared the measured absolute magnitudes of L3--T9 dwarfs with those predicted by the 
precipitating-cloud models of \citeauthor{mar02}\ for varying surface gravities, $g$, and sedimentation 
efficiencies, $f_{\rm sed}$. The models spanning $4.5 \leq {\rm log}~g \leq 5.5$ and $f_{\rm sed} = 3$, 5, and 
``$\infty$'' (no clouds) reproduce well the $M_K$ and $M_{L'}$ of all the dwarfs in our sample. 
The models indicate that the L3--T4.5 dwarfs generally have higher gravities (log~$g = 5.0$--5.5) 
than the T6--T9 dwarfs (log~$g = 4.5$--5.0).  The lower-gravity models underestimate $M_{M'}$ for the 
late-T dwarfs by 0.5--1~mag.  This overestimation of the $M'$ luminosity for $T_{\rm eff} \lesssim 1000$~K is 
attributed to absorption at 4.5--$4.9~\mu$m by CO, which is not expected under the condition of thermochemical 
equilibrium assumed in the models.  The impact of nonequilibrium chemistry on the broadband near-infrared fluxes 
of cool brown dwarfs has only recently been appreciated.  Consequently, the effective-temperature limits of 
space-based $5~\mu$m searches for infra-T dwarfs and EGPs, such as those planned with the recently-launched 
{\it Spitzer Space Telescope}, will be somewhat higher than originally expected.

\acknowledgments The authors thank Didier Saumon for computing the model magnitudes shown in 
Figures~\ref{Golimowski.fig8}--\ref{Golimowski.fig10}.
We also thank the referee, Neill Reid, for many helpful comments.  We gratefully acknowledge the UKIRT staff 
for their assistance with the observations and data acquisition.  Some data were obtained through the UKIRT 
Service Programme.  DAG thanks the Center for Astrophysical Sciences at Johns Hopkins University for its moral
and financial support of this work.  MSM acknowledges support from NASA grants NAG2-6007 and NAG5-8919 and NSF
grant AST~00-86288.   UKIRT is operated by the Joint Astronomy Centre on behalf of the U.~K.\ Particle Physics
and Astronomy Research Council.  Funding for the creation and distribution of the SDSS Archive has been provided
by the Alfred P.\ Sloan Foundation, the Participating Institutions, the National Aeronautics and Space Administration,
the National Science Foundation, the U.~S.\ Department of Energy, the Japanese Monbukagakusho, and the Max Planck 
Society.  The SDSS is managed by the Astrophysical Research Consortium for the Participating Institutions:  The 
University of Chicago, Fermilab, the Institute for Advanced Study, the Japan Participation Group, The Johns 
Hopkins University, Los Alamos National Laboratory, the Max Planck Institute for Astronomy, the Max Planck 
Institute for Astrophysics, New Mexico State University, University of Pittsburgh, Princeton University, the 
United States Naval Observatory, and the University of Washington.  The SDSS Web site is http://www.sdss.org/.

\newpage 

\begin{deluxetable}{llrrccc}
\tabletypesize{\scriptsize}
\tablenum{1}
\tablewidth{420pt}
\tablecaption{The Sample}
\tablehead{
                                & \colhead{Spectral} & \colhead{$\pi$ (error)\tablenotemark{b}} &                                          & \multicolumn{3}{c}{References\tablenotemark{c}}\\
\cline{5-7}
\colhead{Name\tablenotemark{a}} & \colhead{Type}     & \colhead{(mas)}                          & \colhead{$M-m$ (error)\tablenotemark{b}~} & \colhead{SpT} & \colhead{$\pi$} & \colhead{Mult.}
}
\tablecolumns{7}
\startdata 
Gl 229A		           & M1                    & 173.17 (~1.10) &    1.192 (0.014) &  1 &      2,3 &       4 \\
LHS 102A	           & M3.5                  & 104.7~ (11.4~) &    0.100 (0.236) &  5 &        3 &       6 \\
LHS 315		           & M4                    & 298.72 (~1.35) &    2.376 (0.010) &  7 &      2,3 & \nodata \\
LHS 11		           & M4.5                  & 224.8~ (~2.9~) &    1.759 (0.028) &  1 &        3 & \nodata \\
LHS 333AB                  & M5.5 + M7             & 227.9~ (~4.6~) &    1.789 (0.044) &  8 &        3 &       8 \\
LHS 36		           & M6                    & 419.1~ (~2.1~) &    3.112 (0.011) &  9 &        3 & \nodata \\
LHS 292		           & M6.5                  & 220.3~ (~3.6~) &    1.715 (0.035) &  9 &        3 & \nodata \\
LHS 3003	           & M7                    & 156.3~ (~3.0~) &    0.970 (0.042) &  9 &        3 & \nodata \\
LP 326--21\tablenotemark{d}& M8                    & \nodata        &          \nodata & 10 &  \nodata & \nodata \\
LP 349--25\tablenotemark{e}& M8                    & \nodata        &          \nodata & 10 &  \nodata & \nodata \\
LHS 2397aAB                & M8 + L7.5             & ~68.65 (~1.87) & $-$0.817 (0.059) &  8 &     3,11 &      12 \\
TVLM 513--46546	           & M8.5                  & ~94.5~ (~0.6~) & $-$0.123 (0.014) &  9\tablenotemark{f} & 13,14 & \nodata \\ 
LHS 2065		   & M9			   & 117.3~ (~1.5~) &    0.346 (0.028) &  9\tablenotemark{g} &        3 & \nodata \\
LHS 2924	           & M9                    & ~92.4~ (~1.3~) & $-$0.172 (0.031) &  1 &        3 & \nodata \\
BRI 0021--0214	           & M9.5                  & ~84.2~ (~2.6~) & $-$0.373 (0.067) &  9\tablenotemark{h} & 3,13 & \nodata \\
2MASS J03454316+2540233    & L1                    & ~37.1~ (~0.5~) & $-$2.153 (0.029) &  9 &       14 & \nodata \\
2MASS J14392836+1929149    & L1                    & ~69.6~ (~0.5~) & $-$0.787 (0.016) & 15 &       14 & \nodata \\
2MASS J07464256+2000321AB  & L1 + $\sim$L2         & ~81.9~ (~0.3~) & $-$0.434 (0.008) &  9 &       14 &      16 \\
DENIS-P J1058.7--1548      & L3                    & ~57.7~ (~1.0~) & $-$1.194 (0.038) &  9 &       14 & \nodata \\
GD 165B		           & L3                    & ~31.7~ (~2.5~) & $-$2.495 (0.171) &  9 &        3 &      17 \\
Kelu-1		           & L3                    & ~53.6~ (~2.0~) & $-$1.354 (0.081) &  9 &       14 & \nodata \\
2MASS J22244381--0158521   & L3.5                  & ~87.02 (~0.89) & $-$0.302 (0.022) & 15 &    14,18 & \nodata \\
2MASS J00361617+1821104    & L4                    & 114.2~ (~0.8~) &    0.288 (0.015) &  9 &       14 & \nodata \\
LHS 102B	           & L4.5                  & 104.7~ (11.4~) &    0.100 (0.236) & 19 &        3 &       6 \\
SDSS J053951.99--005902.0  & L5                    & ~76.12 (~2.17) & $-$0.593 (0.062) &  9 &       18 & \nodata \\
SDSS J224953.47+004404.6   & L5			   & \nodata        &          \nodata &  9 &  \nodata & \nodata \\
2MASS J15074769--1627386   & L5.5                  & 136.4~ (~0.6~) &    0.674 (0.010) & 15 &       14 & \nodata \\
SDSS J010752.33+004156.1   & L5.5                  & ~64.13 (~4.51) & $-$0.965 (0.153) &  9 &       18 & \nodata \\
DENIS-P J0205.4--1159AB    & L5.5 + L5.5           & ~50.6~ (~1.5~) & $-1.479$ (0.064) &  9 &       14 &      20 \\
2MASS J08251968+2115521    & L6                    & ~94.22 (~0.88) & $-0.129$ (0.020) &  9 &    14,18 & \nodata \\
DENIS-P J1228.2--1547AB    & L6 + $\sim$L6         & ~49.4~ (~1.9~) & $-$1.531 (0.084) &  9 &       14 &      21 \\
2MASS J08503593+1057156AB  & L6 + $\sim$L8.5       & ~33.84 (~2.69) & $-$2.353 (0.173) & 22 &    14,18 &      18 \\
2MASS J16322911+1904407    & L7.5                  & ~65.02 (~1.77) & $-$0.935 (0.059) &  9 &       14 & \nodata \\
2MASS J22443167+2043433    & L7.5                  & \nodata        &          \nodata & 15 &  \nodata & \nodata \\
Gl 584C\tablenotemark{i}   & L8                    & ~54.37 (~1.14) & $-$1.323 (0.046) &  9 &   2,3,18 &      23 \\
SDSS J003259.36+141036.6   & L8                    & ~30.14 (~5.16) & $-$2.604 (0.372) &  9 &       18 & \nodata \\
SDSS J085758.45+570851.4   & L8                    & \nodata        &          \nodata &  9 &  \nodata & \nodata \\
2MASS J03105986+1648155    & L9                    & \nodata        &          \nodata &  9 &  \nodata & \nodata \\
2MASS J09083803+5032088    & L9\tablenotemark{j}   & \nodata        &          \nodata & 15 &  \nodata & \nodata \\
SDSS J080531.80+481233.0   & L9                    & \nodata        &          \nodata & 15 &  \nodata & \nodata \\
SDSS J083008.12+482847.4   & L9                    & ~76.42 (~3.43) & $-$0.584 (0.097) &  9 &       18 & \nodata \\
2MASS J03284265+2302051    & L9.5                  & ~33.13 (~4.20) & $-$2.399 (0.275) &  9 &       18 & \nodata \\
SDSS J204749.61--071818.3  & L9.5                  & \nodata        &          \nodata & 15 &  \nodata & \nodata \\
SDSS J042348.57--041403.5  & T0                    & ~65.93 (~1.70) & $-$0.905 (0.056) &  9 &       18 & \nodata \\
SDSS J120747.17+024424.8   & T0                    & \nodata        &          \nodata & 15 &  \nodata & \nodata \\
SDSS J015141.69+124429.6   & T1                    & ~46.73 (~3.37) & $-$1.652 (0.157) &  9 &       18 & \nodata \\
SDSS J075840.33+324723.4   & T2                    & \nodata        &          \nodata & 15 &  \nodata & \nodata \\
SDSS J125453.90--012247.4  & T2                    & ~73.96 (~1.59) & $-$0.655 (0.047) &  9 &    18,24 & \nodata \\
SDSS J102109.69--030420.1  & T3                    & ~35.35 ( 4.24) & $-$2.258 (0.260) &  9 &    18,24 & \nodata \\
2MASSI J2254188+312349	   & T4                    & \nodata        &          \nodata & 15 &  \nodata & \nodata \\ 
2MASS J05591914--1404488   & T4.5                  & ~96.73 (~0.96) & $-$0.072 (0.022) &  9 &    14,18 & \nodata \\
2MASS J15031961+2525196    & T5.5                  & \nodata        &          \nodata & 25 &  \nodata & \nodata \\
2MASS J15344984--2952274AB & T5.5 + T5.5           & ~73.6~ (~1.2~) & $-$0.666 (0.035) & 26 &       24 &      27 \\
Gl 229B		           & T6                    & 173.17 (~1.10) &    1.192 (0.014) &  9 &      2,3 &       4 \\
2MASSI J0243137--245329    & T6                    & ~93.62 (~3.63) & $-$0.143 (0.084) & 26 &       18 & \nodata \\
2MASS J09373487+2931409	   & T6\tablenotemark{k}   & 162.84 (~3.88) &    1.059 (0.052) & 15 &       18 & \nodata \\
SDSS J123147.39+084730.7   & T6                    & \nodata        &          \nodata & 15 &  \nodata & \nodata \\
SDSS J162414.37+002915.6   & T6                    & ~90.73 (~1.03) & $-$0.211 (0.025) &  9 & 14,18,24 & \nodata \\
2MASS J12255432--2739466AB & T6 + T8               & ~74.79 (~2.03) & $-$0.631 (0.059) &  9 &    18,24 &      27 \\
Gl 570D\tablenotemark{l}   & T8                    & 170.16 (~1.45) &    1.154 (0.019) &  9 &      2,3 &      28 \\
2MASSI J0727182+171001     & T8                    & 110.14 (~2.34) &    0.210 (0.046) & 15 &       18 & \nodata \\
2MASS J12171110--0311131   & T8                    &  93.20 (~2.06) & $-$0.153 (0.048) &  9 &    18,24 & \nodata \\
2MASSI J0415195--093506    & T9                    & 174.34 (~2.76) &    1.207 (0.034) & 15 &  \nodata & \nodata \\
\enddata

\tablenotetext{a}{\scriptsize IAU approved designations for 2MASS and SDSS point sources are 
``2MASS Jhhmmss[.]ss$\pm$ddmmss[.]s'' and ``SDSS Jhhmmss.ss$\pm$ddmmss.s,'' where 
the equatorial coordinates are given at equinox J2000.  Preliminary designations
are given for 2MASS sources whose IAU-approved designations are unpublished.}
\tablenotetext{b}{\scriptsize Based on weighted mean of referenced trigonometric parallaxes.}
\tablenotetext{c}{\scriptsize References for principal spectral type, trigonometric parallax, and multiplicity: 
(1) \citealt{kir91},
(2) \citealt{esa97},
(3) \citealt{van95},
(4) \citealt{nak95},
(5) \citealt{mar99b},
(6) \citealt{gol99},
(7) \citealt{hen02},
(8) \citealt{gli91},
(9) G02,
(10) \citealt{giz00},
(11) \citealt{tin96},
(12) \citealt{fre03},
(13) \citealt{tin95},
(14) \citealt{dah02},
(15) \citealt{kna04},
(16) \citealt{rei01},
(17) \citealt{bec88},
(18) \citealt{vrb04},
(19) \citealt{leg02b},
(20) \citealt{koe99},
(21) \citealt{mar99a},
(22) \citealt{kir99b},
(23) \citealt{kir01},
(24) \citealt{tin03},
(25) \citealt{bur03b},
(26) Classified on scheme of G02 using spectra of \citealt{bur02a},
(27) \citealt{bur03c}
(28) \citealt{bur00}
} 
\tablenotetext{d}{\scriptsize Also known as 2MASSW J1444171+300214.}
\tablenotetext{e}{\scriptsize Also known as 2MASSW J0027559+221932.}
\tablenotetext{f}{\scriptsize Name abbreviated to T~513 by G02.}
\tablenotetext{g}{\scriptsize Misidentified as ``LHS 2025'' by G02.}
\tablenotetext{h}{\scriptsize Name abbreviated to BRI~0021 by G02.}
\tablenotetext{i}{\scriptsize Also known as 2MASS J15232263+3014562.}
\tablenotetext{j}{\scriptsize Classified as L5 by \citet{cru03} from optical spectrum.}
\tablenotetext{k}{\scriptsize Labelled ``peculiar'' by \citet{bur02a} because of low $K$-band flux.}
\tablenotetext{l}{\scriptsize Also known as 2MASSW J1457150--212148.} 

\end{deluxetable}

\newpage

\begin{deluxetable}{lclcclc}
\tabletypesize{\footnotesize}
\tablenum{2}
\tablecaption{New MKO $L'$ and $M'$ Photometry}
\tablehead{
\colhead{Name} & \colhead{$L'$ (error)} & \colhead{Imager} & \colhead{Date} & 
\colhead{$M'$ (error)} & \colhead{Imager} & \colhead{Date} \\
}
\tablecolumns{7}
\startdata 
SDSS J0032+1410     & 13.35 (0.05) & UIST    & 2003 Sep 04 & \nodata      & \nodata & \nodata     \\
SDSS J0151+1244     & 13.54 (0.05) & IRCAM   & 2001 Nov 25 & \nodata      & \nodata & \nodata     \\
DENIS J0205--1159AB & \nodata      & \nodata & \nodata     & 12.10 (0.20) & IRCAM   & 2001 Nov 23 \\
2MASS J0243--2453   & 13.25 (0.05) & IRCAM   & 2001 Nov 24 & \nodata      & \nodata & \nodata     \\
2MASS J0328+2302    & 13.33 (0.05) & UIST    & 2003 Nov 08 & \nodata      & \nodata & \nodata     \\
2MASS J0415--0935   & 13.28 (0.05) & IRCAM   & 2001 Nov 25 & 12.82 (0.15) & IRCAM   & 2001 Nov 25 \\
SDSS J0423--0414    & \nodata      & \nodata & \nodata     & 11.90 (0.05) & UIST    & 2003 Jan 04 \\
SDSS J0539--0059    & \nodata      & \nodata & \nodata     & 11.87 (0.10) & UIST    & 2003 Jan 04 \\
2MASS J0727+1710    & 13.68 (0.05) & IRCAM   & 2001 Nov 25 & \nodata      & \nodata & \nodata     \\
SDSS J0758+3247     & 11.94 (0.03) & UIST    & 2003 Jan 04 & \nodata      & \nodata & \nodata     \\
SDSS J0758+3247     & 12.06 (0.05) & UIST    & 2003 Nov 10 & \nodata      & \nodata & \nodata     \\
SDSS J0805+4812     & 12.31 (0.05) & UIST    & 2003 Nov 10 & \nodata      & \nodata & \nodata     \\
2MASS J0908+5032    & 11.37 (0.06) & IRCAM   & 2002 Jun 18 & 11.95 (0.20) & UIST    & 2002 Dec 06 \\
2MASS J0937+2931    & 12.34 (0.05) & IRCAM   & 2001 Nov 24 & 11.74 (0.10) & IRCAM   & 2001 Nov 25 \\
Gl 229A             & \nodata      & \nodata & \nodata     & ~4.04 (0.05) & IRCAM   & 2001 Nov 23 \\
Gl 229B             & 12.24 (0.05) & IRCAM   & 2001 Nov 23 & 11.74 (0.10) & IRCAM   & 2001 Nov 23 \\
SDSS J1021--0304    & 13.64 (0.05) & IRCAM   & 2001 Nov 23 & \nodata      & \nodata & \nodata     \\
SDSS J1207+0244     & 12.62 (0.05) & UIST    & 2003 May 16 & \nodata      & \nodata & \nodata     \\
SDSS J1231+0847     & 13.52 (0.05) & UIST    & 2003 May 16 & \nodata      & \nodata & \nodata     \\
Kelu-1              & \nodata      & \nodata & \nodata     & 11.22 (0.10) & IRCAM   & 2002 Jun 21 \\
2MASS J1439+1929    & \nodata      & \nodata & \nodata     & 11.13 (0.06) & IRCAM   & 2002 Jun 18 \\
2MASS J1503+2525    & 11.91 (0.05) & UIST    & 2003 Jan 04 & 12.25 (0.15) & UIST    & 2003 May 16 \\
2MASS J1507--1627   & \nodata      & \nodata & \nodata     & 10.69 (0.05) & IRCAM   & 2002 Jun 18 \\
2MASS J1534--2952AB & 12.58 (0.05) & UIST    & 2003 May 17 & \nodata      & \nodata & \nodata     \\
SDSS J2047--0718    & 13.80 (0.05) & IRCAM   & 2001 Nov 25 & \nodata      & \nodata & \nodata     \\
2MASS J2224--0158   & 10.90 (0.05) & IRCAM   & 2002 Jul 15 & 11.32 (0.05) & UIST    & 2003 Jun 04 \\
2MASS J2244+2043    & 12.11 (0.03) & UIST    & 2003 Jan 04 & \nodata      & \nodata & \nodata     \\
SDSS J2249+0044     & 12.71 (0.07) & UIST    & 2003 Jun 18 & \nodata      & \nodata & \nodata     \\
2MASS J2254+3123    & 13.24 (0.05) & IRCAM   & 2001 Nov 24 & \nodata      & \nodata & \nodata     \\

\enddata

\end{deluxetable}

\newpage

\begin{deluxetable}{llrrrr}
\tabletypesize{\footnotesize}
\tablenum{3}
\tablecaption{MKO $KL'M'$ Photometry of M, L, and T Dwarfs}
\tablehead{
\colhead{Name} & \colhead{Spectral Type} & \colhead{$M_{L'}$ (error)\tablenotemark{a}} & \colhead{$L'$ (error)} &
\colhead{$K$--$L'$ (error)} & \colhead{$L'$--$M'$ (error)}
}
\tablecolumns{6}
\startdata 
Gl 229A\tablenotemark{b}             & M1            &  5.25 (0.05) &  4.06 (0.05) & 0.09 (0.07) &  0.02 (0.07)\\ 
LHS 102A\tablenotemark{c}            & M3.5          &  7.63 (0.24) &  7.53 (0.05) & 0.20 (0.06) & \nodata \\
LHS 315                              & M4            &  7.62 (0.01) &  5.24 (0.01) & 0.41 (0.04) & \nodata \\
LHS 11                               & M4.5          &  8.08 (0.08) &  6.32 (0.07) & 0.33 (0.08) & \nodata \\
LHS 333AB                            & M5.5 + M7     &  7.42 (0.07) &  5.63 (0.05) & 0.45 (0.06) & \nodata \\
LHS 36\tablenotemark{d}              & M6            &  8.82 (0.05) &  5.71 (0.05) & 0.35 (0.06) & $-$0.14 (0.06) \\
LHS 292\tablenotemark{d}             & M6.5          &  9.16 (0.06) &  7.45 (0.05) & 0.50 (0.06) & $-$0.20 (0.07) \\
LHS 3003                             & M7            &  9.40 (0.05) &  8.43 (0.03) & 0.50 (0.05) & \nodata \\
LP 326--21\tablenotemark{e}	     & M8            &      \nodata & 10.09 (0.07) &     \nodata & $-$0.30 (0.12) \\
LP 349--25\tablenotemark{e}	     & M8            &      \nodata &  9.15 (0.07) &     \nodata & $-$0.24 (0.12) \\
LHS 2397aAB                          & M8 + L7.5     &  9.21 (0.06) & 10.03 (0.02) & 0.66 (0.04) & \nodata \\
TVLM 513--46546                      & M8.5          &  9.92 (0.08) & 10.04 (0.08) & 0.65 (0.09) & \nodata \\
LHS 2065\tablenotemark{e,{\rm f}}    & M9            &  9.74 (0.08) &  9.39 (0.07) & 0.52 (0.05) & $-$0.23 (0.10) \\
LHS 2924                             & M9            &  9.95 (0.04) & 10.12 (0.03) & 0.60 (0.05) & \nodata \\
BRI 0021--0214                       & M9.5          &  9.41 (0.15) &  9.78 (0.13) & 0.75 (0.14) & \nodata \\
2MASS J0345+2540                     & L1            &  9.86 (0.10) & 12.01 (0.10) & 0.65 (0.11) & \nodata \\
2MASS J1439+1929                     & L1            & 10.01 (0.05) & 10.80 (0.05) & 0.67 (0.06) & $-$0.33 (0.08) \\
2MASS J0746+2000AB\tablenotemark{d}  & L1 + $\sim$L2 &  9.24 (0.03) &  9.67 (0.03) & 0.76 (0.04) & $-$0.35 (0.08) \\
DENIS J1058--1548                    & L3            & 10.43 (0.11) & 11.62 (0.10) & 0.93 (0.11) & \nodata \\
GD 165B                              & L3            & 10.43 (0.18) & 12.93 (0.07) & 1.16 (0.08) & \nodata \\
Kelu-1                               & L3            &  9.43 (0.17) & 10.78 (0.15) & 1.00 (0.16) & $-$0.44 (0.18) \\
2MASS J2224--0158\tablenotemark{g}   & L3.5          & 10.60 (0.05) & 10.90 (0.05) & 1.08 (0.06) & $-$0.42 (0.07) \\
2MASS J0036+1821                     & L4            & 10.37 (0.05) & 10.08 (0.05) & 0.96 (0.06) & $-$0.27 (0.07) \\
LHS 102B\tablenotemark{c}            & L4.5          & 10.51 (0.24) & 10.41 (0.05) & 0.95 (0.06) & \nodata \\
SDSS J0539--0059                     & L5            & 10.73 (0.08) & 11.32 (0.05) & 1.08 (0.06) & $-$0.55 (0.11) \\
SDSS J2249+00                        & L5            & \nodata	    & 12.71 (0.07) & 1.69 (0.08) & \nodata \\
2MASS J1507--1627                    & L5.5          & 10.65 (0.03) &  9.98 (0.03) & 1.31 (0.04) & $-$0.71 (0.06) \\
SDSS J0107+0041                      & L5.5          & 11.10 (0.17) & 12.06 (0.07) & 1.52 (0.08) & \nodata \\
DENIS J0205--1159AB                  & L5.5 + L5.5   &  9.96 (0.12) & 11.44 (0.10) & 1.55 (0.10) & $-$0.66 (0.22) \\
2MASS J0825+2115                     & L6            & 11.40 (0.04) & 11.53 (0.03) & 1.40 (0.04) & \nodata \\
DENIS J1228--1547AB                  & L6 + $\sim$L6 &  9.89 (0.13) & 11.42 (0.10) & 1.29 (0.11) & \nodata \\
2MASS J0850+1057AB                   & L6 + $\sim$L8.5 & 10.59 (0.18) & 12.94 (0.05) & 1.41 (0.06) & \nodata \\
2MASS J1632+1904                     & L7.5          & 11.60 (0.08) & 12.54 (0.05) & 1.43 (0.07) & \nodata \\
2MASS J2244+2043\tablenotemark{g}    & L7.5          & \nodata      & 12.11 (0.03) & 1.79 (0.04) & \nodata \\
Gl 584C                              & L8            & 11.54 (0.07) & 12.86 (0.05) & 1.49 (0.07) & \nodata \\
SDSS J0032+1410                      & L8            & 10.75 (0.38) & 13.35 (0.05) & 1.64 (0.07) & \nodata \\
SDSS J0857+5708                      & L8            & \nodata      & 11.31 (0.05) & 1.63 (0.06) & $-$0.19 (0.11) \\
2MASS J0310+1648                     & L9            & \nodata      & 12.54 (0.05) & 1.64 (0.06) & \nodata \\
2MASS J0908+5032\tablenotemark{g}    & L9            & \nodata      & 11.37 (0.06) & 1.52 (0.07) & $-$0.58 (0.21) \\
SDSS J0805+4812                      & L9            & \nodata      & 12.31 (0.05) & 1.20 (0.06) & \nodata \\
SDSS J0830+4828                      & L9            & 11.40 (0.11) & 11.98 (0.05) & 1.70 (0.06) & \nodata \\
2MASS J0328+2302		     & L9.5          & 10.93 (0.28) & 13.33 (0.05) & 1.54 (0.06) & \nodata \\
SDSS J2047--0718\tablenotemark{g}    & L9.5          & \nodata      & 13.80 (0.05) & 1.54 (0.06) & \nodata \\
SDSS J0423--0414                     & T0            & 10.55 (0.08) & 11.45 (0.05) & 1.51 (0.06) & $-$0.45 (0.07) \\
SDSS J1207+0244\tablenotemark{g}     & T0            & \nodata      & 12.62 (0.05) & 1.54 (0.06) & \nodata \\
SDSS J0151+1244                      & T1            & 11.89 (0.16) & 13.54 (0.05) & 1.64 (0.07) & \nodata \\
SDSS J0758+3247\tablenotemark{g,{\rm h}}& T2            & \nodata      & 11.97 (0.03) & 1.90 (0.04) & \nodata \\
SDSS J1254--0122                     & T2            & 11.60 (0.07) & 12.25 (0.05) & 1.59 (0.06) & $-$0.40 (0.21) \\
SDSS J1021--0304                     & T3            & 11.38 (0.26) & 13.64 (0.05) & 1.62 (0.07) & \nodata \\
2MASS J2254+3123\tablenotemark{g}    & T4            & \nodata      & 13.24 (0.05) & 1.79 (0.06) & \nodata \\
2MASS J0559--1404                    & T4.5          & 12.07 (0.05) & 12.14 (0.05) & 1.59 (0.06) & $-$0.01 (0.16) \\
2MASS J1503+2525\tablenotemark{g}    & T5.5          & \nodata      & 11.91 (0.05) & 2.08 (0.06) & $-$0.34 (0.16) \\
2MASS J1534--2952AB\tablenotemark{g} & T5.5 + T5.5   & 11.91 (0.06) & 12.58 (0.05) & 2.33 (0.06) & \nodata \\
Gl 229B                              & T6            & 13.43 (0.05) & 12.24 (0.05) & 2.12 (0.06) & 0.50 (0.11) \\
2MASS J0243--2453\tablenotemark{g}   & T6            & 13.11 (0.10) & 13.25 (0.05) & 2.09 (0.06) & \nodata \\
2MASS J0937+2931\tablenotemark{g}    & T6            & 13.40 (0.07) & 12.34 (0.05) & 3.05 (0.08) & 0.60 (0.11) \\
SDSS J1231+0847\tablenotemark{g}     & T6            & \nodata      & 13.52 (0.05) & 1.94 (0.06) & \nodata \\
SDSS J1624+0029                      & T6            & 13.39 (0.05) & 13.60 (0.04) & 2.01 (0.06) & \nodata \\
2MASS J1225--2739AB                  & T6 + T8       & 12.59 (0.10) & 13.22 (0.08) & 2.06 (0.09) & \nodata \\
Gl 570D                              & T8            & 14.13 (0.05) & 12.98 (0.05) & 2.54 (0.07) & \nodata \\
2MASS J0727+1710\tablenotemark{g}    & T8            & 13.89 (0.07) & 13.68 (0.05) & 2.01 (0.06) & \nodata \\
2MASS J1217--0311                    & T8            & 13.81 (0.07) & 13.96 (0.05) & 1.96 (0.06) & \nodata \\
2MASS J0415--0935\tablenotemark{g}   & T9            & 14.49 (0.06) & 13.28 (0.05) & 2.55 (0.06) & 0.46 (0.16) \\
\enddata

\tablenotetext{a}{Based on weighted mean trigonometric parallaxes.  See Table~1.}
\tablenotetext{b}{$L'$ from \citet{leg02b}.}
\tablenotetext{c}{$KL'$ from \citet{leg02b}.}
\tablenotetext{d}{$M'$ from \citet{rei02}.}
\tablenotetext{e}{$L'M'$ from \citet{rei02}.}
\tablenotetext{f}{$K$ synthesized from spectra of G02.}
\tablenotetext{g}{$K$ from \citet{kna04}.}
\tablenotetext{h}{$L'$ is weighted mean of values listed in Table~2.}
\end{deluxetable}

\clearpage

\newpage 

\begin{deluxetable}{cccrrrrccc}
\tabletypesize{\scriptsize}
\rotate
\tablenum{4}
\tablewidth{570pt}
\tablecaption{Polynomial Fits to Diagrams}
\tablehead{
& & & \multicolumn{7}{c}{Polynomial coefficients} \\
\cline{4-10} 
\colhead{$P(x)$\tablenotemark{a}} & \colhead{$x$} & \colhead{RMS\tablenotemark{b}} & \colhead{$c_0$} & \colhead{$c_1$} & \colhead{$c_2$} & \colhead{$c_3$} & \colhead{$c_4$} & \colhead{$c_5$} & \colhead{$c_6$}
}
\tablecolumns{10}
\startdata
$M_K$    & $0.09 \leq K$--$L' \leq 2.55$ & 0.49 & 4.0760e+00 & 1.9467e+01 & $-$2.1584e+01   & 1.1235e+01   & $-$1.9583e+00 & \nodata & \nodata  \\
$M_K$    & $0.21 \leq K$--$M' \leq 3.01$ & 0.43 & 8.2327e+00 & 6.9722e+00 & $-$3.3255e+00   & 6.5907e$-$01 & \nodata & \nodata & \nodata\\
$M_{L'}$ & M1 $\leq$ SpT\tablenotemark{c}~ $\leq$ T9 & 0.35 & 4.3095e+00 & 1.1450e+00 & $-$8.0385e$-$02 & 2.4832e$-$03 & $-$2.2539e$-$05 & \nodata & \nodata\\
$M_{M'}$ & M1 $\leq$ SpT\tablenotemark{c}~ $\leq$ T9 & 0.45 & 3.5211e+00 & 1.1826e+00 & $-$6.4508e$-$02 & 1.2549e$-$03 & \nodata & \nodata & \nodata\\
BC$_K$ & M6 $\leq$ SpT\tablenotemark{c}~ $\leq$ T9 & 0.13 & 3.9257e+00 & $-$3.8338e$-$01 & 5.3597e$-$02 & $-$2.6550e$-$03 & 4.0859e$-$05 & \nodata & \nodata \\
$T_{\rm eff}$ & M6 $\leq$ SpT\tablenotemark{c}~ $\leq$ T9 & 124~K & 1.4322e+04 & $-$5.1287e+03 & 9.0951e+02 & $-$8.3099e+01 & 4.0323e+00 & $-$9.8598e$-$02 & 9.5373e$-$04 \\
\enddata

\tablenotetext{a}{$P(x) = \displaystyle{\sum_{i=0}^{n}} c_{\scriptscriptstyle i} x^{\scriptscriptstyle i}$}
\tablenotetext{b}{Units are magnitudes except where noted.}
\tablenotetext{c}{Fit requires numerical translation of spectral types: M1--M9.5 $\rightarrow$ 1--9.5, 
L0--L9.5 $\rightarrow$ 10--19.5, T0--T9 $\rightarrow$ 20--29.}
\end{deluxetable}

\newpage
\begin{deluxetable}{llrrrrccc}
\rotate
\tabletypesize{\footnotesize}
\tablenum{5}
\tablewidth{540pt}
\tablecaption{Supplemental Dwarfs Lacking $L'M'$ Measurements}
\tablehead{
                          & \colhead{Spectral} & \colhead{$\pi$ (error)\tablenotemark{b}} &  & \colhead {Measured} & \colhead{Estimated\tablenotemark{c}} & \multicolumn{3}{c}{References\tablenotemark{d}}\\
\cline{7-9}
\colhead{Name\tablenotemark{a}} & \colhead{Type}     & \colhead{(mas)} & \colhead{$M-m$ (error)\tablenotemark{b}~} & \colhead{$M_K$ (error)} & \colhead{$M_{L'}$ (error)} & \colhead{SpT} & \colhead{$\pi$} & \colhead{$K$}
}
\tablecolumns{10}
\startdata
SDSS J225529.09-003433.4  & M8.5 & 16.19 (~2.59) & $-$3.954 (0.347) & 10.33 (0.35) &  9.68 (0.36) & 1 & 2 & 3 \\
SDSS J144600.60+002452.0  & L5   & 45.46 (~3.25) & $-$1.712 (0.155) & 12.09 (0.16) & 10.83 (0.23) & 1 & 2 & 3 \\
SDSS J132629.82--003831.5 & L5.5 & 49.98 (~6.33) & $-$1.506 (0.275) & 12.66 (0.28) & 11.25 (0.32) & 1 & 2 & 4 \\
SDSS J083717.21--000018.0 & T0.5 & 33.70 (13.45) & $-$2.362 (0.867) & 13.62 (0.87) & 12.04 (0.88) & 1 & 2 & 3 \\
SDSS J175032.96+175903.9  & T3.5 & 36.24 (~4.53) & $-$2.204 (0.271) & 13.82 (0.28) & 12.11 (0.32) & 1 & 2 & 3 \\
SDSS J020742.83+000056.2  & T4.5 & 34.85 (~9.87) & $-$2.289 (0.615) & 14.33 (0.62) & 12.54 (0.64) & 1 & 2 & 3 \\
2MASSI J2356547--155310   & T6   & 68.97 (~3.42) & $-$0.807 (0.108) & 14.92 (0.11) & 12.98 (0.19) & 5 & 2 & 4 \\
SDSS J134646.45-003150.4  & T6   & 69.07 (~2.09) & $-$0.804 (0.066) & 14.93 (0.08) & 12.87 (0.18) & 1 & 2,6 & 3 \\
2MASSI J1047538+212423    & T6.5 & 98.75 (~3.30) & $-$0.027 (0.073) & 16.17 (0.08) & 13.67 (0.18) & 1 & 2,6 & 3 \\
\enddata
\tablenotetext{a}{Naming protocol as described in Table~1.}
\tablenotetext{b}{Based on weighted mean of referenced trigonometric parallaxes.}
\tablenotetext{c}{$L'$ estimated from dwarfs in Table~1 with similar spectral types and $JHK$ colors.
Errors include dispersions from fits to $K$--$L'$ versus spectral type (\S4.1) of 0.06~mag (M dwarfs), 0.15~mag 
(L dwarfs), and 0.16~mag (T dwarfs).}
\tablenotetext{d}{References for spectral type, trigonometric parallax, and $K$ photometry: 
(1) G02,
(2) V04,
(3) L02,
(4) K04,
(5) Classified on scheme of G02 using spectra of \citealt{bur02a},
(6) \citealt{tin03}
}
\end{deluxetable}

\newpage
\begin{deluxetable}{llrccrr}
\tabletypesize{\scriptsize}
\tablenum{6}
\tablewidth{455pt}
\tablecaption{Bolometric Luminosity and Effective Temperature}
\tablehead{
               &                         &                                &                           &                                                 & \multicolumn{2}{c}{$T_{\rm eff}$ (K)} \\
\cline{6-7}
\colhead{Name} &  \colhead{Spectral Type}&\colhead{$M_{\rm bol}$ (error)} &  \colhead{BC$_K$ (error)} &  \colhead{log$(L_{\rm bol}/L_{\odot})$ (error)} & \colhead{Range\tablenotemark{a}} & \colhead{3~Gyr\tablenotemark{b}}
}
\tablecolumns{7}
\startdata 
Gl 229A\tablenotemark{c}          & M1             &  7.97 (0.09) & 2.63 (0.07) & $-$1.29 (0.02) & 3750--3775 & 3755 \\
LHS 102A\tablenotemark{c}         & M3.5           & 10.55 (0.25) & 2.72 (0.06) & $-$2.32 (0.10) & 3200--3300 & 3275 \\
LHS 36                            & M6             & 12.18 (0.08) & 3.01 (0.07) & $-$2.97 (0.02) & 2650--2900 & 2900 \\
LHS 292                           & M6.5           & 12.65 (0.09) & 2.98 (0.07) & $-$3.16 (0.03) & 2475--2750 & 2725 \\
LHS 3003                          & M7             & 12.95 (0.09) & 3.05 (0.07) & $-$3.28 (0.03) & 2350--2650 & 2600 \\
SDSS J2255--0034\tablenotemark{d} & M8.5           & 13.51 (0.36) & 3.18 (0.07) & $-$3.50 (0.14) & 2000--2525 & 2400 \\
TVLM 513--46546                   & M8.5           & 13.73 (0.08) & 3.16 (0.07) & $-$3.59 (0.02) & 2025--2325 & 2300 \\
LHS 2065                          & M9             & 13.47 (0.09) & 3.21 (0.07) & $-$3.49 (0.02) & 2150--2425 & 2400 \\
BRI 0021--0214                    & M9.5           & 13.37 (0.10) & 3.21 (0.07) & $-$3.45 (0.03) & 2150--2475 & 2425 \\
2MASS J0345+2540                  & L1             & 13.75 (0.08) & 3.24 (0.07) & $-$3.60 (0.02) & 2000--2325 & 2300 \\
2MASS J1439+1929                  & L1             & 13.88 (0.07) & 3.20 (0.06) & $-$3.66 (0.02) & 1950--2275 & 2250 \\
2MASS J0746+2000AB                & L1 + $\sim$L2  & 13.26 (0.07) & 3.26 (0.06) & $-$3.41 (0.02) & 1900--2225\tablenotemark{e} & 2200\tablenotemark{e}\\
DENIS J1058--1548                 & L3             & 14.73 (0.09) & 3.37 (0.07) & $-$4.00 (0.03) & 1600--1950 & 1900 \\
GD 165B                           & L3             & 14.90 (0.19) & 3.31 (0.07) & $-$4.06 (0.07) & 1750--1925 & 1850 \\
Kelu-1                            & L3             & 13.74 (0.11) & 3.31 (0.07) & $-$3.59 (0.04) & 2100--2350 & 2300\tablenotemark{f} \\
2MASS J2224--0158                 & L3.5           & 15.14 (0.07) & 3.46 (0.06) & $-$4.15 (0.02) & 1475--1800 & 1750 \\
2MASS J0036+1821                  & L4             & 14.67 (0.07) & 3.34 (0.06) & $-$3.97 (0.02) & 1650--1975 & 1900 \\
LHS 102B\tablenotemark{c}         & L4.5           & 14.89 (0.25) & 3.43 (0.06) & $-$4.05 (0.10) & 1750--1975 & 1850 \\
SDSS J0539--0059                  & L5             & 15.12 (0.09) & 3.31 (0.06) & $-$4.15 (0.03) & 1475--1800 & 1750 \\
SDSS J1446+0024\tablenotemark{d}  & L5             & 15.43 (0.18) & 3.34 (0.07) & $-$4.27 (0.07) & 1300--1725 & 1650 \\
2MASS J1507--1627                 & L5.5           & 15.16 (0.07) & 3.20 (0.06) & $-$4.16 (0.02) & 1475--1800 & 1750 \\
SDSS J0107+0041                   & L5.5           & 15.93 (0.17) & 3.32 (0.06) & $-$4.47 (0.06) & 1175--1550 & 1475 \\
SDSS J1326--0038\tablenotemark{d} & L5.5           & 15.94 (0.28) & 3.28 (0.06) & $-$4.48 (0.11) & 1150--1600 & 1475 \\
DENIS J0205--1159AB               & L5.5 + L5.5    & 14.71 (0.09) & 3.20 (0.06) & $-$3.98 (0.03) & 1350--1700\tablenotemark{e} & 1650\tablenotemark{e}\\
2MASS J0825+2115                  & L6             & 16.10 (0.07) & 3.30 (0.06) & $-$4.54 (0.02) & 1175--1475 & 1425 \\
DENIS J1228--1547AB               & L6 + L6        & 14.50 (0.11) & 3.32 (0.07) & $-$3.90 (0.04) & 1400--1775\tablenotemark{e} & 1700\tablenotemark{e} \\
2MASS J1632+1904                  & L7.5           & 16.23 (0.11) & 3.19 (0.07) & $-$4.59 (0.03) & 1150--1450 & 1375 \\
Gl 584C                           & L8             & 16.20 (0.11) & 3.17 (0.09) & $-$4.58 (0.04) & 1300--1400 & 1350\tablenotemark{f} \\
SDSS J0032+1410\tablenotemark{d}  & L8             & 15.46 (0.39) & 3.07 (0.09) & $-$4.28 (0.15) & 1250--1800 & 1650 \\
SDSS J0830+4828                   & L9             & 16.19 (0.13) & 3.09 (0.08) & $-$4.58 (0.05) & 1125--1475 & 1400 \\
2MASS J0328+2302\tablenotemark{d} & L9.5           & 15.53 (0.29) & 3.06 (0.08) & $-$4.31 (0.11) & 1250--1750 & 1625 \\
SDSS J0423--0414                  & T0             & 15.11 (0.10) & 3.05 (0.08) & $-$4.14 (0.04) & 1450--1825 & 1750 \\
SDSS J0837--0000\tablenotemark{d} & T0.5           & 16.50 (0.87) & 2.88 (0.09) & $-$4.70 (0.35) &  900--1600 & 1300 \\
SDSS J0151+1244                   & T1             & 16.46 (0.19) & 2.93 (0.09) & $-$4.68 (0.07) & 1050--1425 & 1300 \\
SDSS J1254--0122                  & T2             & 16.08 (0.10) & 2.90 (0.08) & $-$4.54 (0.04) & 1150--1500 & 1425 \\
SDSS J1021--0304                  & T3             & 15.76 (0.28) & 2.76 (0.09) & $-$4.40 (0.11) & 1200--1650 & 1525 \\
SDSS J1750+1759\tablenotemark{d}  & T3.5           & 16.35 (0.29) & 2.53 (0.09) & $-$4.64 (0.11) & 1050--1475 & 1350 \\
2MASS J0559--1404                 & T4.5           & 16.07 (0.13) & 2.41 (0.13) & $-$4.53 (0.05) & 1150--1500 & 1425 \\
SDSS J0207+0000\tablenotemark{d}  & T4.5           & 16.80 (0.63) & 2.47 (0.13) & $-$4.82 (0.25) &  875--1450 & 1200 \\
2MASS J0243--2453                 & T6             & 17.45 (0.15) & 2.25 (0.13) & $-$5.08 (0.06) &  825--1150 & 1025 \\
2MASS J0937+2931                  & T6             & 17.96 (0.16) & 1.51 (0.14)\tablenotemark{g} & $-$5.28 (0.05) & 725--1000 & 900 \\
2MASS J2356--1553\tablenotemark{d}& T6             & 17.26 (0.17) & 2.34 (0.13) & $-$5.00 (0.06) &  875--1200 & 1075 \\
Gl 229B\tablenotemark{c}          & T6             & 17.77 (0.08) & 2.22 (0.07) & $-$5.21 (0.02) &  850--1050 &  950 \\
SDSS J1346--0031\tablenotemark{d} & T6             & 17.25 (0.15) & 2.32 (0.13) & $-$5.00 (0.06) &  875--1200 & 1075 \\
SDSS J1624+0029                   & T6             & 17.64 (0.14) & 2.24 (0.13) & $-$5.16 (0.05) &  800--1100 &  975 \\
2MASS J1225--2739AB               & T6 + T8        & 16.86 (0.14) & 2.21 (0.13) & $-$4.85 (0.05) &  800--1100\tablenotemark{e} & 975\tablenotemark{e}\\
2MASS J1047+2124		  & T6.5           & 18.13 (0.15) & 1.96 (0.13) & $-$5.35 (0.06) & 725--950 &  900 \\ 
2MASS J0727+1710                  & T8             & 18.14 (0.14) & 2.24 (0.13) & $-$5.35 (0.05) & 725--950 &  900 \\ 
2MASS J1217--0311                 & T8             & 18.05 (0.14) & 2.28 (0.13) & $-$5.32 (0.05) & 725--975 &  900 \\ 
Gl 570D                           & T8             & 18.57 (0.14) & 1.90 (0.13) & $-$5.53 (0.05)\tablenotemark{h} & 784--824\tablenotemark{h} &       800 \\ 
2MASS J0415--0935                 & T9             & 19.07 (0.13) & 2.03 (0.13) & $-$5.73 (0.05) & 600--750 &  700 \\ 
\enddata

\tablenotetext{a}{Range of $T_{\rm eff}$ for assumed ages of 0.1--10~Gyr and known parallax uncertainties.  The ages of Gl~229AB, LHS~102AB, 
GD~165B, Gl~584C, Gl~570D, and Kelu-1 have been further constrained observationally (see footnote~13).}
\tablenotetext{b}{$T_{\rm eff}$ at age $\sim 3$~Gyr, unless otherwise noted.}
\tablenotetext{c}{$M_{\rm bol}$, BC$_K$ and log$(L_{\rm bol}/L_{\odot})$ from \citet{leg02b}.}
\tablenotetext{d}{$L'$ estimated from spectral type and $JHK$ colors (\S4.3).}
\tablenotetext{e}{Assuming uneclipsed components of equal luminosity.}
\tablenotetext{f}{$T_{\rm eff}$ given for middle of age range given in footnote 13.}
\tablenotetext{g}{Strongly depressed $K$-band flux produces atypical BC$_K$.}
\tablenotetext{h}{log$(L_{\rm bol}/L_{\odot})$ and $T_{\rm eff}$ from \citet{geb01}.}
\end{deluxetable}

\newpage
\begin{figure}[t]
   \epsscale{1}\plotone{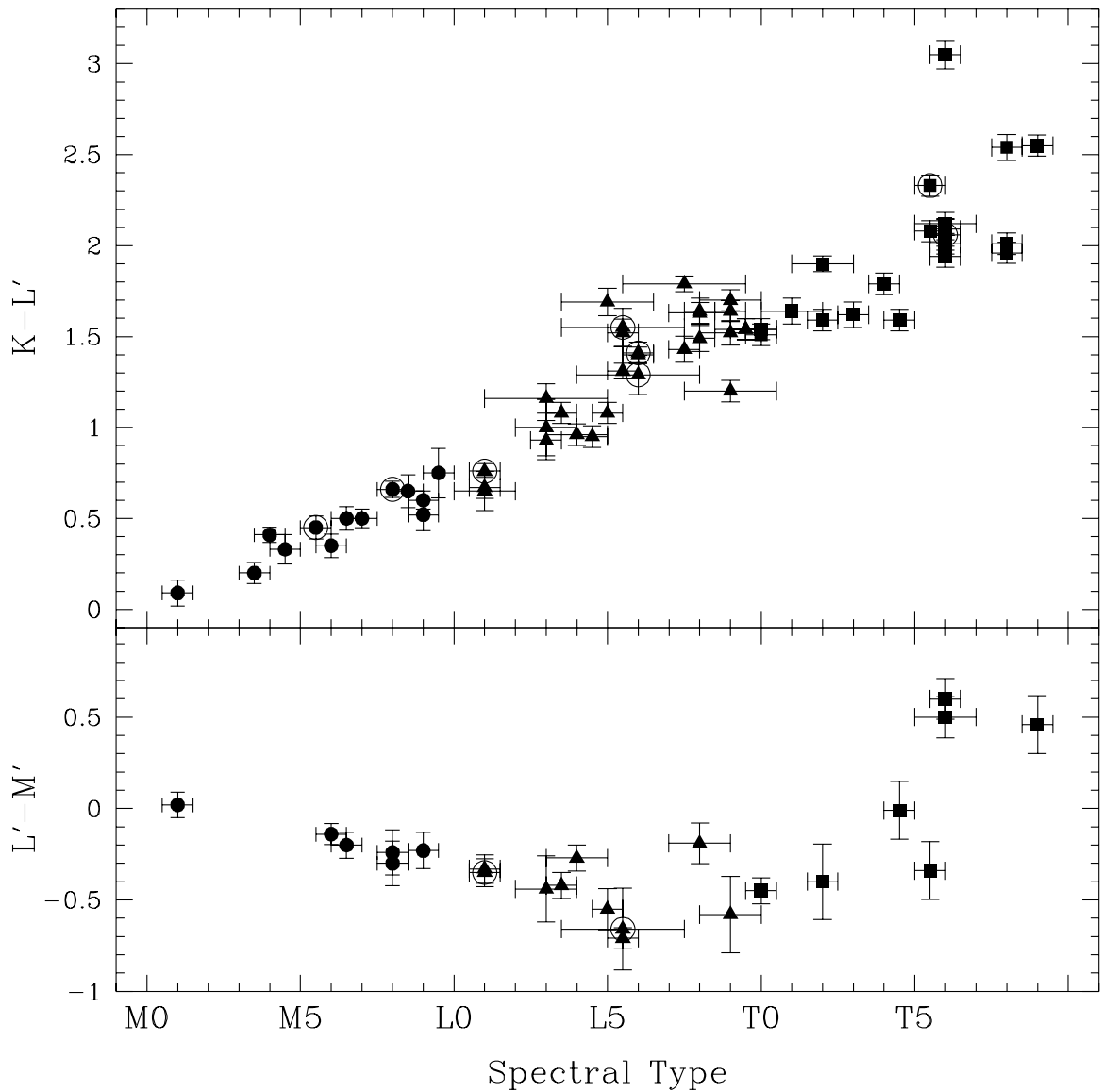}
   \caption{
      Variations of $K$--$L'$ and $L'$--$M'$ with spectral type for M dwarfs (circles), L dwarfs 
      (triangles), and T dwarfs (squares) listed in Table~3.  
      All data are based on the MKO photometric system.  
      Points representing close-binary systems are surrounded by open circles.  
      The L9 dwarf with the anomalously blue $K$--$L' = 1.20$ is SDSS~J0805+4812, and 
      the T6 dwarf with the anomalously red $K$--$L' = 3.05$ is 2MASS~J0937+2931.  Both dwarfs
      are discussed in \S4.5.
   }
  \label{Golimowski.fig1}
\end{figure}

\newpage
\begin{figure}[t]
   \epsscale{1}\plotone{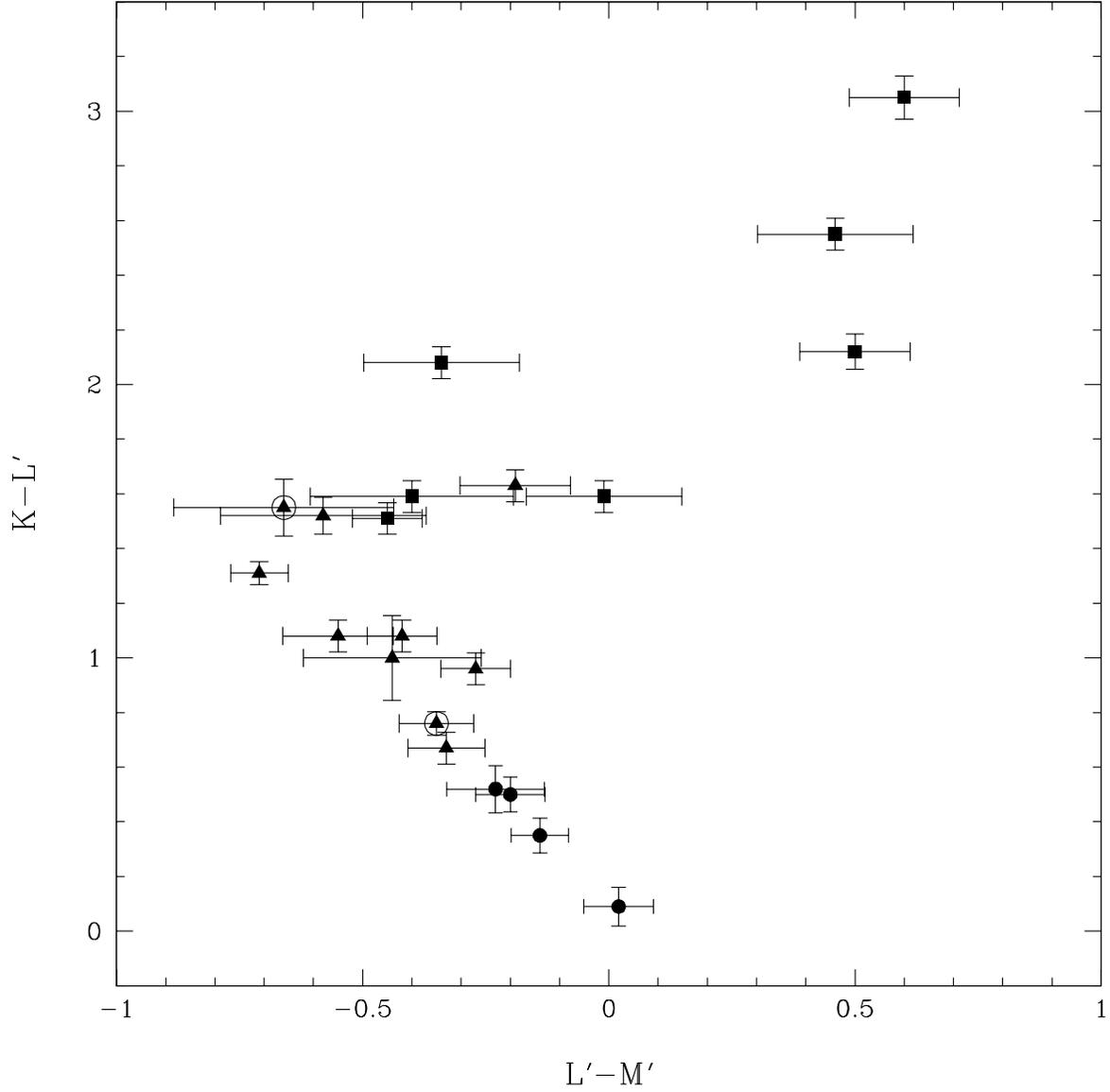}
   \caption{
      Color--color diagram of $K$--$L'$ versus $L'$--$M'$ for M, L, and T dwarfs listed in Table~3.
      All data are based on the MKO photometric system.
      All symbols are described in Figure~\ref{Golimowski.fig1}.
   }
  \label{Golimowski.fig2}
\end{figure}

\newpage
\begin{figure}[t]
   \epsscale{1}\plotone{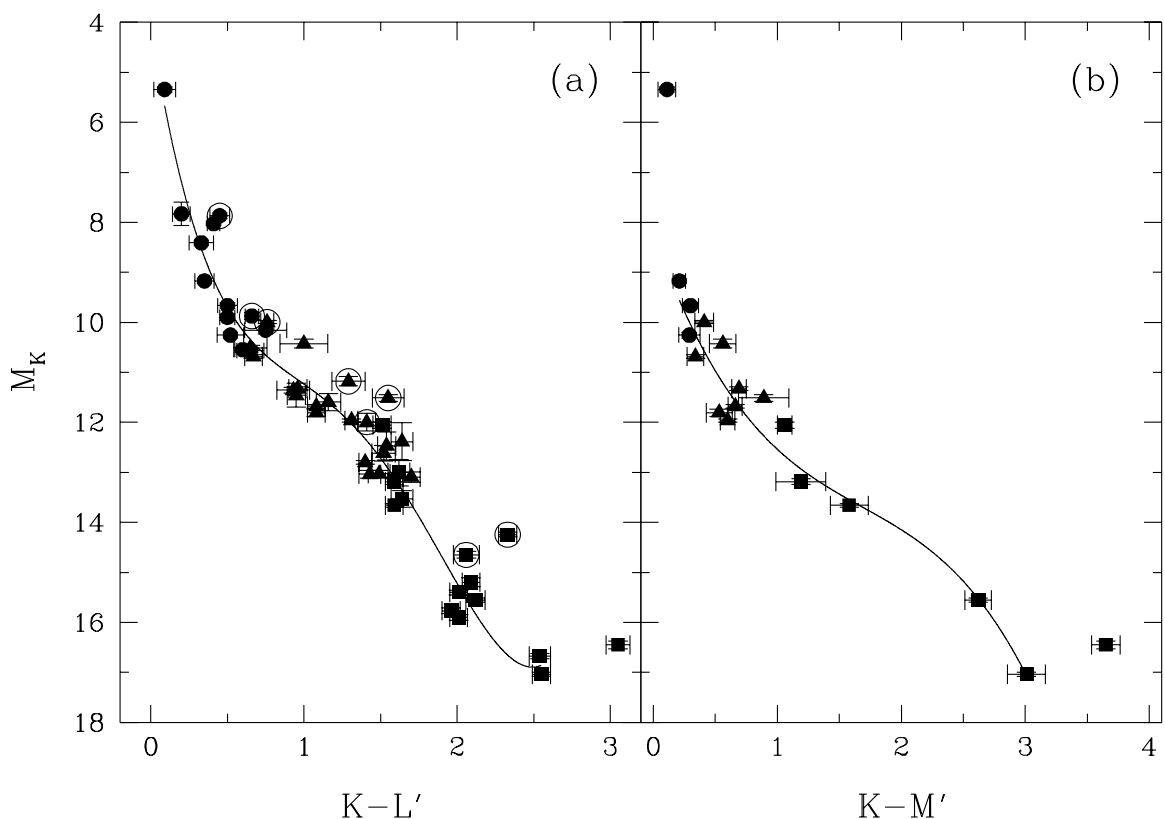}
   \caption{
      Color--magnitude diagrams of (a) $M_K$ versus $K$--$L'$ and (b) $M_K$ versus 
      $K$--$M'$ for M, L, and T dwarfs listed in Table~3.
      All data are based on the MKO photometric system.
      All symbols are described in Figure~\ref{Golimowski.fig1}.
      The $M_K$ extrema represent Gl~229A (M1) and 2MASS~J0415--0935 (T9).
      The curves are (a) fourth-order and (b) third-order polynomial fits to the unweighted
      data except those representing known close-binary systems (encircled points) and the 
      anomalously red 2MASS~J0937+2931 ($K$--$L' = 3.05$, $K$--$M' = 3.65$).  The datum for 
      Gl~229A ($K$--$M' = 0.11$) was also omitted from the fit in (b).
   }
  \label{Golimowski.fig3}
\end{figure}

\newpage
\begin{figure}[t]
   \epsscale{1}\plotone{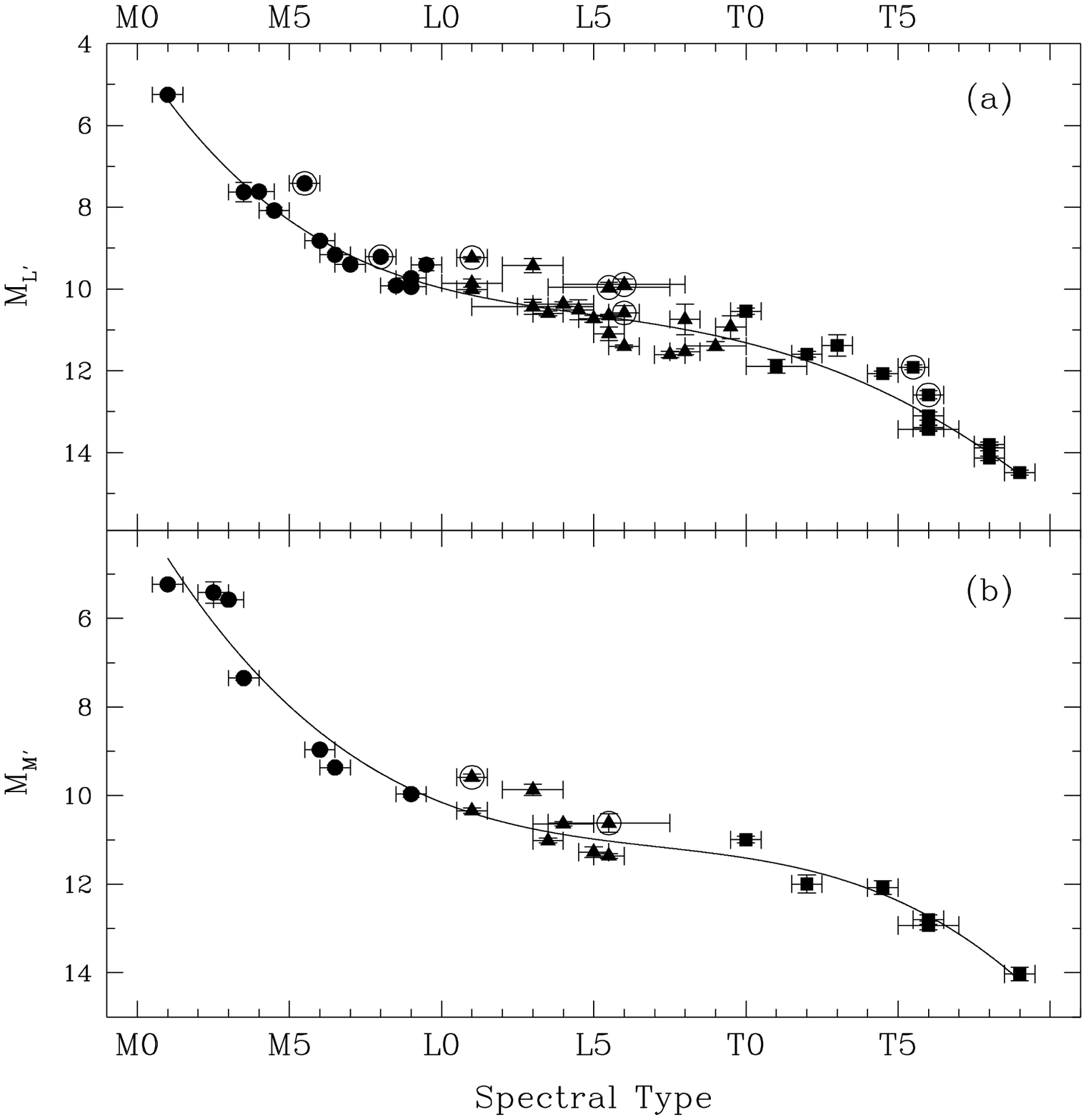}
   \caption{
      Diagrams of (a) $M_{L'}$ and (b) $M_{M'}$ versus spectral type for M, L, and T dwarfs listed 
      in Table~3.  
      All data are based on the MKO photometric system.
      All symbols are described in Figure~\ref{Golimowski.fig1}.
      Diagram (b) is supplemented with MKO $M'$ measurements reported by \citet{rei02} for Gl~811.1 
      (M2.5; G02), Gl~752A \citep[M3;][]{kir91}, and Gl~643 (M3.5; G02).  
      The weighted means of the parallaxes of these M dwarfs measured by Yale Observatory \citep{van95}
      and {\it Hipparcos} \citep{esa97} are, respectively, $55.81 \pm 6.27$~mas,
      $171.01 \pm 0.62$~mas, and $158.28 \pm 3.45$~mas.
      The curves are (a) fourth-order and (b) third-order polynomial fits to the unweighted
      data except those representing known close-binary systems (encircled points).
   }
  \label{Golimowski.fig4}
\end{figure}

\newpage
\begin{figure}[t]
   \epsscale{1}\plotone{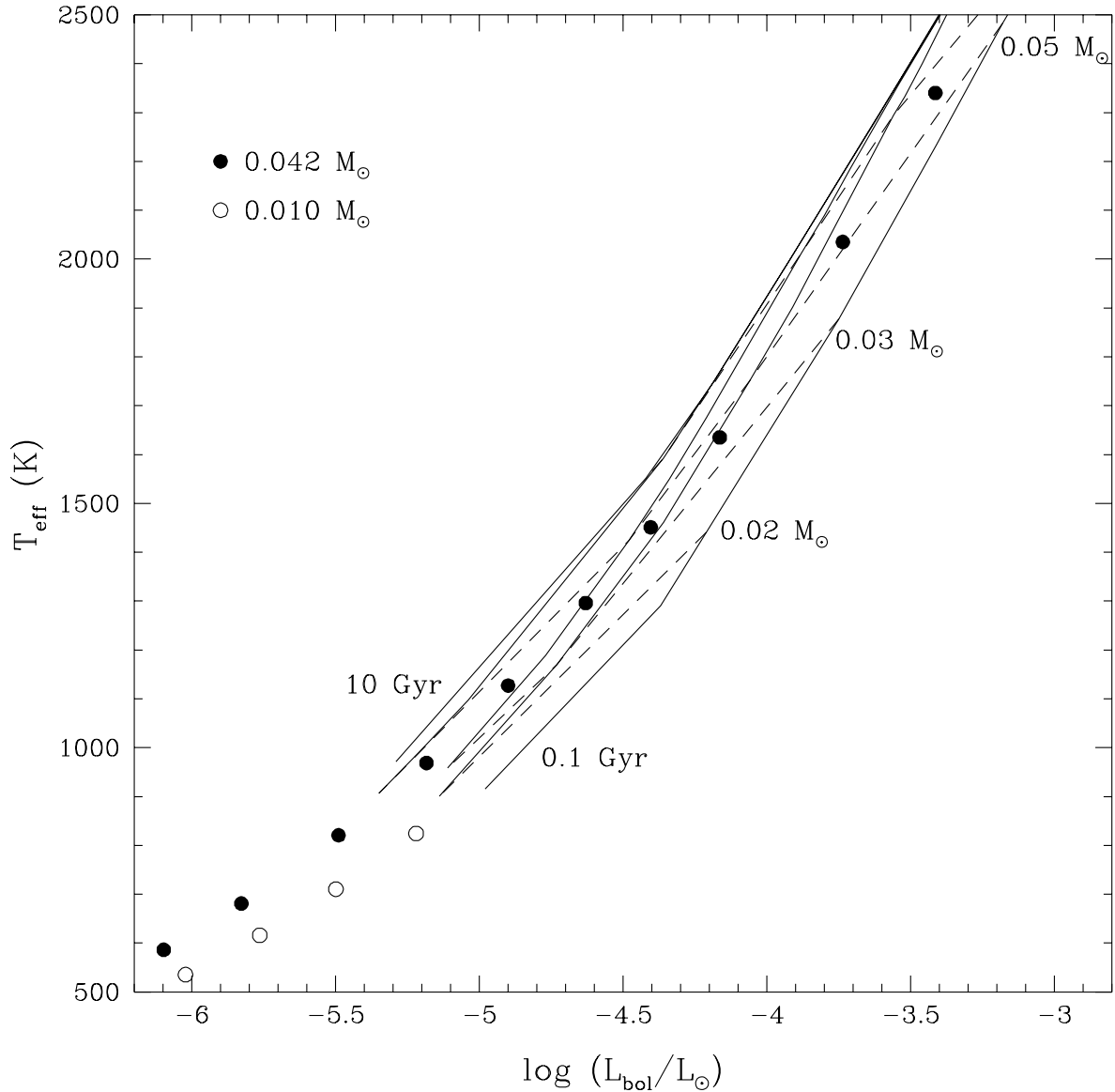}
   \caption{\footnotesize{
    Predicted evolutions of $L_{\rm bol}$ and $T_{\rm eff}$ as functions of mass and age.  The diagram is an 
    extension to lower $T_{\rm eff}$ of Figure~12 of \citet{leg01}.  The solid curves are, from right to left, the 
    0.1, 0.5, 1, 5, and 10~Gyr isochrones for 0.01--$0.08~M_{\odot}$ brown dwarfs computed from the DUSTY 
    atmosphere models of \citet{cha00}.  The dashed curves are, from top to bottom, the cooling tracks for 
    0.07, 0.05, 0.03, and $0.02~M_{\odot}$ brown dwarfs computed from the same models.  Also shown are the 
    cooling tracks for $0.042~M_{\odot}$ (filled circles) and $0.010~M_{\odot}$ (open circles) brown dwarfs of 
    ages 0.1--10~Gyr and 0.1--0.5~Gyr, respectively, computed from the settled-dust models of \citet{bur97} 
    for time intervals of $\sim 0.2$~dex.  Despite the differences between the two models' treatment of 
    photospheric condensates, the predicted cooling tracks from each model are mutually consistent.  The 
    range of $T_{\rm eff}$ for fixed $L_{\rm bol}$ never exceeds $\sim 300$~K.}
   }
  \label{Golimowski.fig5}
\end{figure}

\newpage
\begin{figure}[t]
   \epsscale{1}\plotone{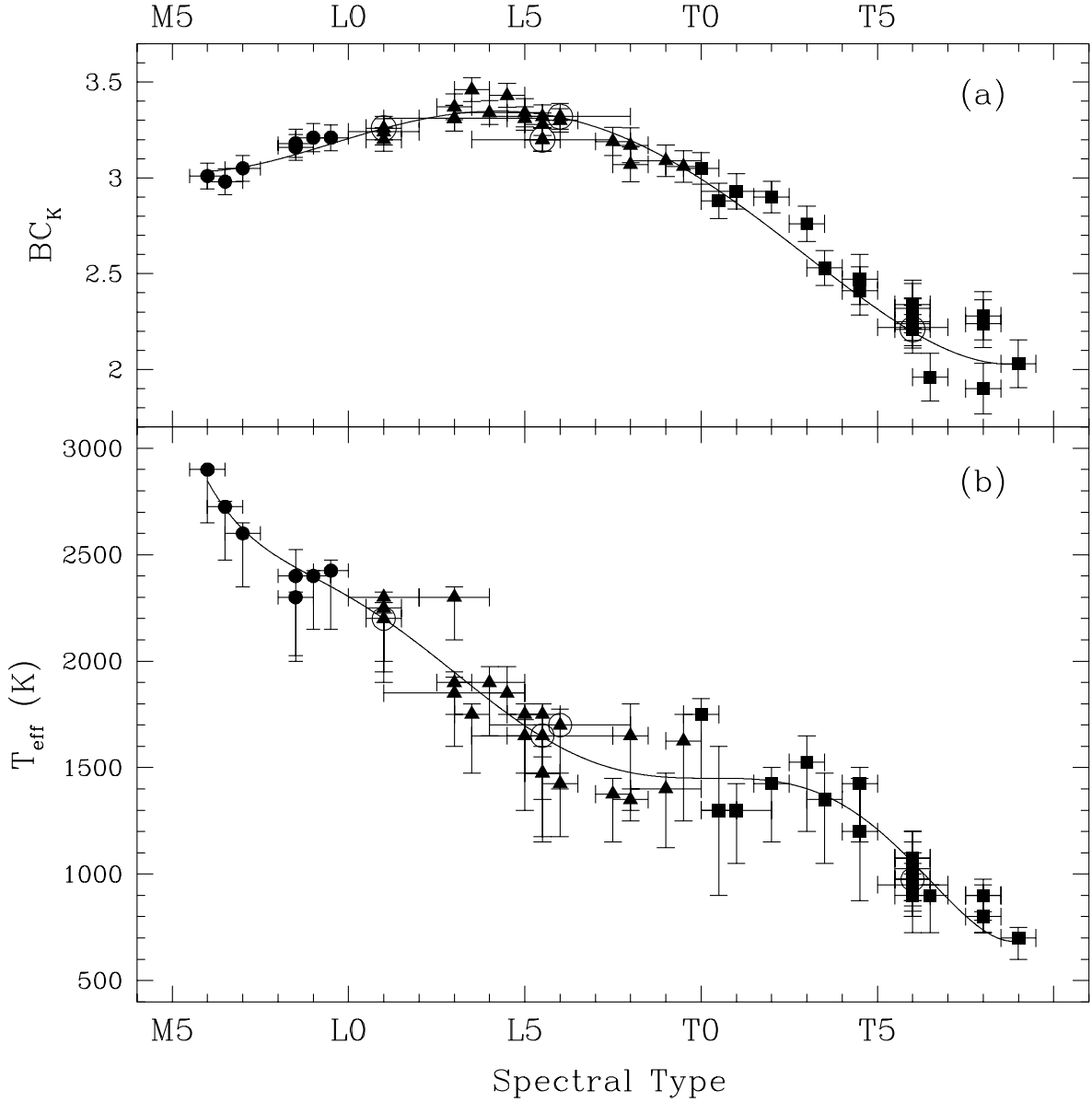}
   \caption{
      Diagrams of (a) BC$_K$ and (b) $T_{\rm eff}$ versus spectral type for ultracool dwarfs
      listed in Table~6.  All symbols are described in Figure~\ref{Golimowski.fig1}.  The plotted values 
      of $T_{\rm eff}$ are those listed in Column~7 of Table~6 for a mean age of 3~Gyr, unless 
      otherwise noted.  The error bars for these values reflect the full ranges of $T_{\rm eff}$ 
      listed in Column~6 of Table~6.
      The curves are (a) fourth-order and (b) sixth-order polynomial fits to the weighted data
      except those representing known close-binary systems (encircled points).  The datum for the
      T6 dwarf 2MASS~J0937+2931 (BC$_K = 1.51$) is not shown in (a), but it is included in the
      polynomial fit.  The fit in (b) is fixed at type T9 to avoid an unrealistic upturn in
      $T_{\rm eff}$ between types T8 and T9.
   }
  \label{Golimowski.fig6}
\end{figure}

\newpage
\begin{figure}[t]
   \epsscale{1}\plotone{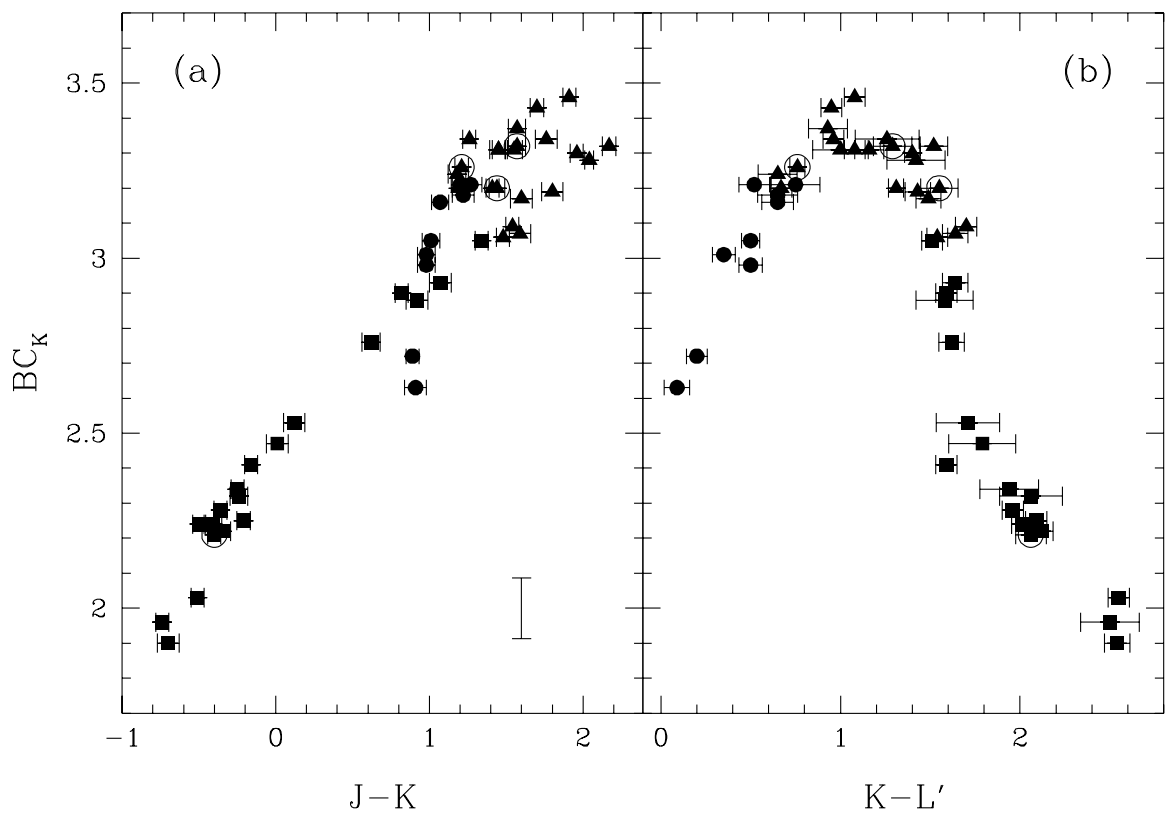}
   \caption{
      Diagrams of BC$_K$ versus (a) $J$--$K$ and (b) $K$--$L'$ for dwarfs
      listed in Table~6.  
      All data are based on the MKO photometric system;
      $J$ measurements are taken from L02 and K04.
      All symbols are described in Figure~\ref{Golimowski.fig1}.
      The uncertainty in BC$_K$ for each point is omitted for clarity; the average uncertainty 
      is represented by the vertical error bar in the lower right corner of (a).
      The data for the T6 dwarf 2MASS~J0937+2931 (BC$_K = 1.51$) are not shown.
   }
  \label{Golimowski.fig7}
\end{figure}

\newpage
\begin{figure}[t]
   \epsscale{1}\plotone{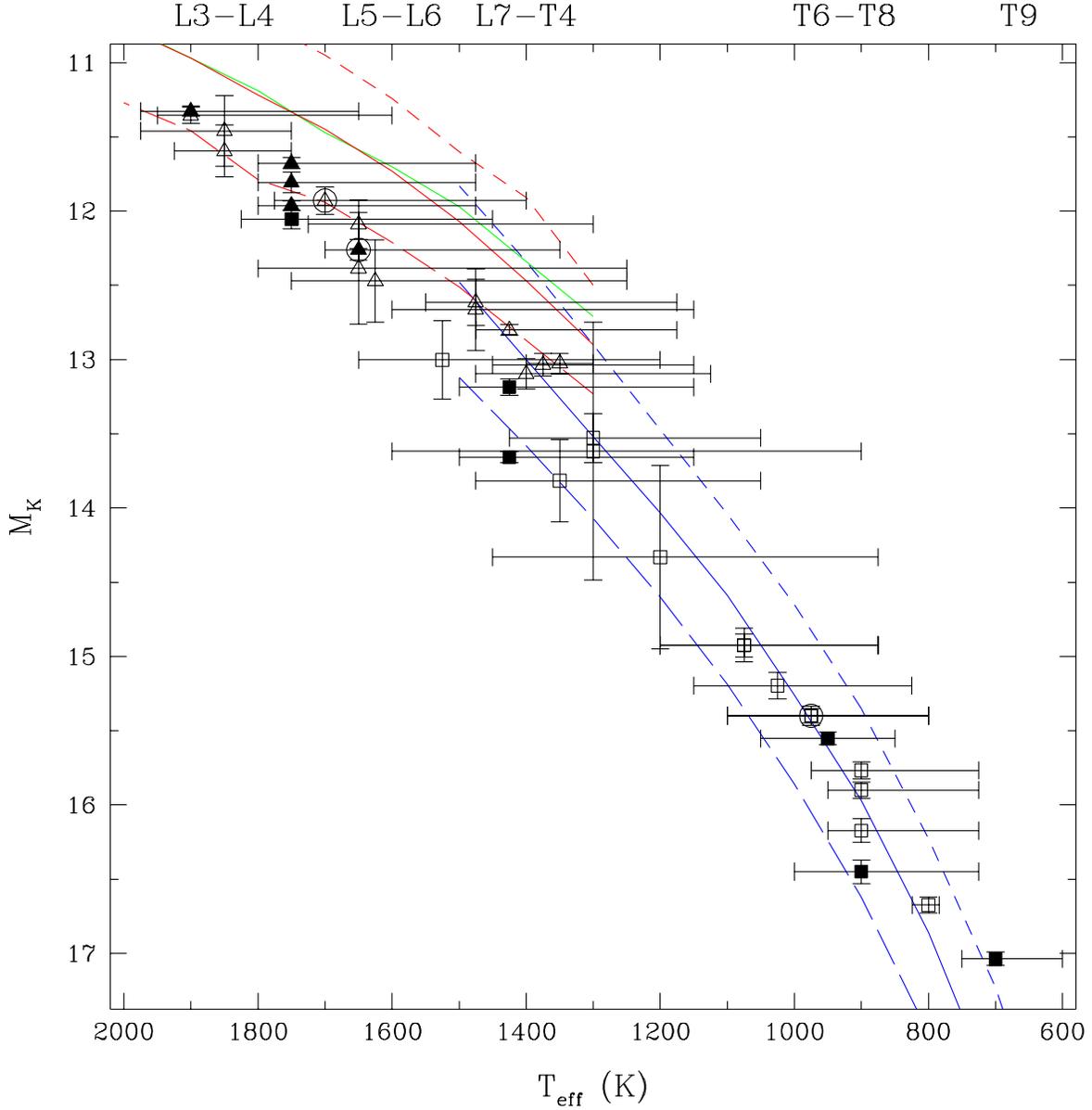}
   \caption{
      Diagram of $M_K$ versus $T_{\rm eff}$ for L3--T9 dwarfs listed in Table~6.  
      The plotted values of $T_{\rm eff}$ are those listed in Column~7 of Table~6 for a mean age of
      $\sim 3$~Gyr, unless otherwise noted.  The error bars for these values reflect the full ranges of 
      $T_{\rm eff}$ listed in Column~6 of Table~6.  L dwarfs are 
      represented by triangles and T dwarfs are represented by squares.  Filled symbols denote 
      those dwarfs for which we have $M'$ photometric data (Figure~\ref{Golimowski.fig10}).
      The measured values of $M_K$ for close binaries (encircled points) have been increased by 
      0.75~mag to represent one component of the presumed uneclipsed, equal-luminosity systems.
      The curves are the predicted relationships from the models of \citet{mar02} for brown dwarfs
      with $f_{\rm sed} = 3$ (green), $f_{\rm sed} = 5$ (red), and cloud-free atmospheres (blue),
      and surface gravities of log~$g = 4.5$ (short-dash), 5.0 (solid), and 5.5 (long-dash).
   }
  \label{Golimowski.fig8}
\end{figure}

\newpage
\begin{figure}[t]
   \epsscale{1}\plotone{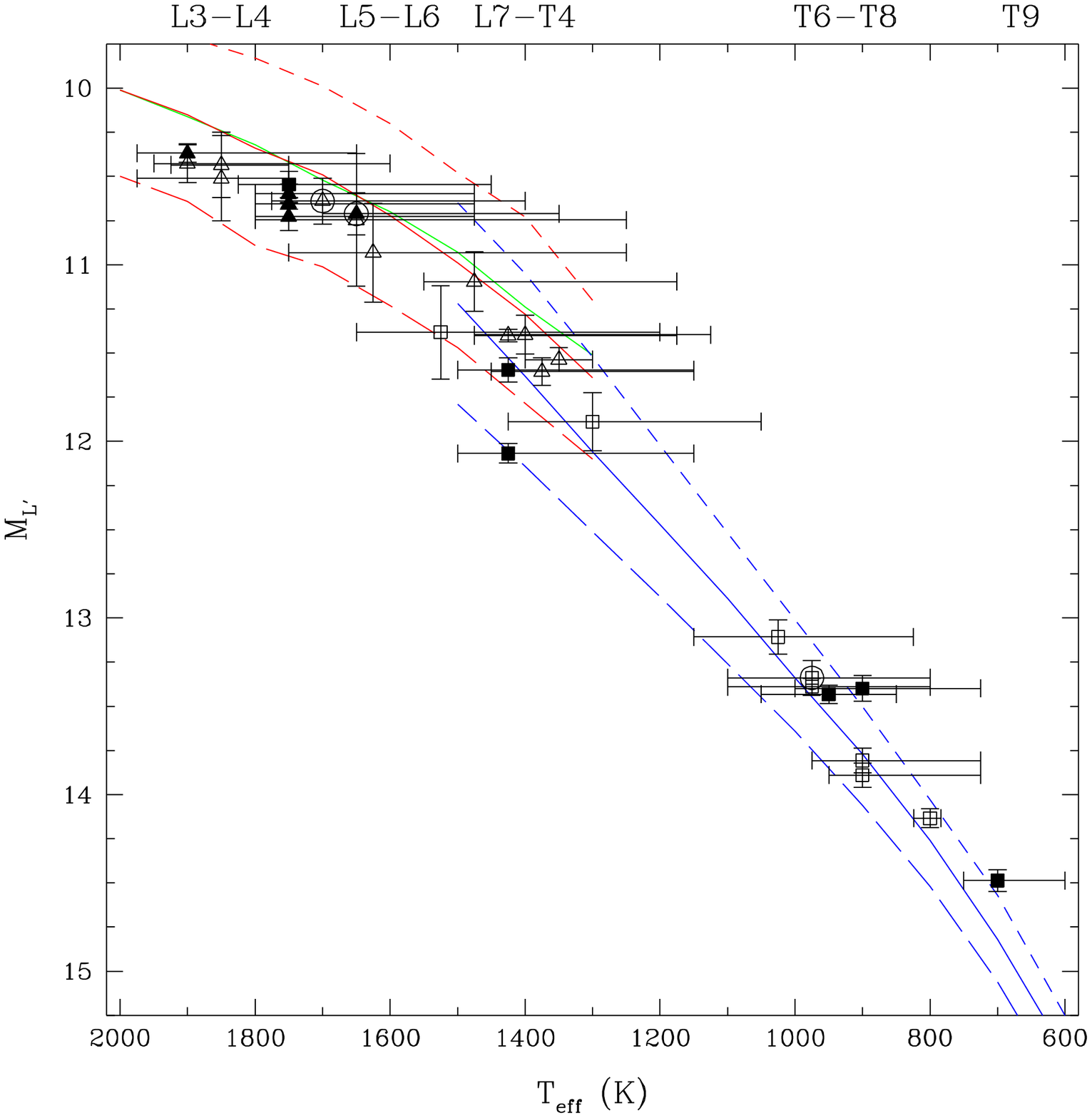}
   \caption{
      Diagram of $M_{L'}$ versus $T_{\rm eff}$ for L3--T9 dwarfs listed in Table~6.  
      All symbols and curves are described in Figure~\ref{Golimowski.fig8}.  
   }
  \label{Golimowski.fig9}
\end{figure}

\newpage
\begin{figure}[t]
   \epsscale{1}\plotone{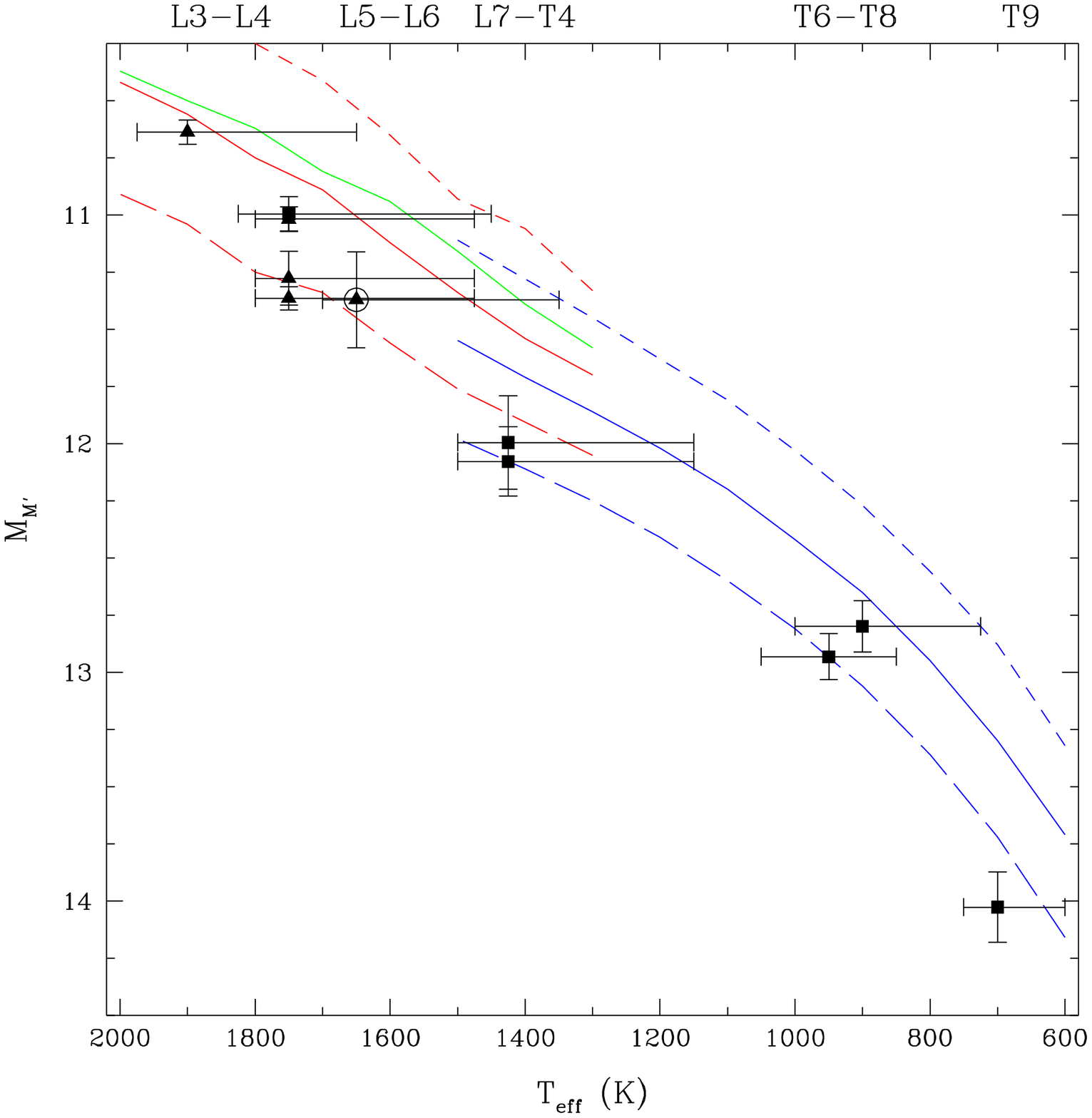}
   \caption{
      Diagram of $M_{M'}$ versus $T_{\rm eff}$ for L3--T9 dwarfs listed in Table~6. 
      All symbols and curves are described in Figure~\ref{Golimowski.fig8}.  
   }
  \label{Golimowski.fig10}
\end{figure}

\end{document}